\documentclass[11pt]{article}%
\usepackage{amsmath}
\usepackage{graphicx}
\usepackage{amsfonts}
\usepackage{amssymb}%
\newtheorem{theorem}{Theorem}

\newtheorem{conjecture}[theorem]{Conjecture}
\newtheorem{corollary}[theorem]{Corollary}

\newtheorem{definition}[theorem]{Definition}

\newtheorem{lemma}[theorem]{Lemma}

\newtheorem{proposition}[theorem]{Proposition}

\newenvironment{proof}[1][Proof]{\textbf{#1.} }{\ \rule{0.5em}{0.5em}}
\begin{document}

\title{Multilinear Formulas and Skepticism of Quantum Computing}
\author{Scott Aaronson\thanks{University of California, Berkeley. \ Email:
aaronson@cs.berkeley.edu. \ Part of this work was done at the Perimeter
Institute (Waterloo, Canada). \ Supported by an NSF Graduate Fellowship and by
the Defense Advanced Research Projects Agency (DARPA).}}
\date{}
\maketitle

\begin{abstract}
Several researchers, including Leonid Levin, Gerard 't Hooft, and Stephen
Wolfram, have argued that quantum mechanics will break down before the
factoring of large numbers becomes possible. \ If this is true, then there
should be a natural set of quantum states that can account for all quantum
computing experiments performed to date, but \textit{not} for Shor's factoring
algorithm. \ We investigate as a candidate the set of states expressible by a
polynomial number of additions and tensor products. \ Using a recent lower
bound on multilinear formula size due to Raz, we then show that states arising
in quantum error-correction require $n^{\Omega(\log n)}$ additions and tensor
products even to approximate, which incidentally yields the first
superpolynomial gap between general and multilinear formula size of functions.
\ More broadly, we introduce a complexity classification of pure quantum
states, and prove many basic facts about this classification. \ Our goal is to
refine vague ideas about a breakdown of quantum mechanics into specific
hypotheses that might be experimentally testable in the near future.

\end{abstract}

\section{Introduction\label{INTRO}}

\begin{quote}
QC of the sort that factors long numbers seems firmly rooted in science
fiction \ldots\ The present attitude would be analogous to, say, Maxwell
selling the Daemon of his famous thought experiment as a path to cheaper
electricity from heat. ---Leonid Levin \cite{levin}
\end{quote}

Quantum computing presents a dilemma:\ is it reasonable to study a type of
computer that has never been built, and might never be built in one's
lifetime? \ Some researchers strongly believe the answer is `no.' \ Their
objections generally fall into four categories:

\begin{enumerate}
\item[(A)] There is a fundamental physical reason why large quantum computers
can never be built.

\item[(B)] Even if (A) fails, large quantum computers will never be built in practice.

\item[(C)] Even if (A) and (B) fail, the speedup offered by quantum computers
is of limited theoretical interest.

\item[(D)] Even if (A), (B), and (C) fail, the speedup is of limited practical
value.\footnote{Because of the `even if' clauses, the objections seem to us
logically independent, so that there are $16$ possible positions regarding
them (or $15$ if one is against quantum computing). \ We ignore the
possibility that no speedup exists, in other words that $\mathsf{BPP}%
=\mathsf{BQP}$. \ By `large quantum computer' we mean any computer much faster
than its best classical simulation, as a result of asymptotic complexity
rather than the speed of elementary operations. \ Such a computer need not be
universal; it might be specialized for (say) factoring.}
\end{enumerate}

The objections can be classified along two axes:%
\[%
\begin{tabular}
[c]{lll}
& \textbf{Theoretical} & \textbf{Practical}\\
\textbf{Physical} & (A) & (B)\\
\textbf{Algorithmic} & (C) & (D)
\end{tabular}
\
\]
This paper focuses on objection (A). \ Its goal is not to win a debate about
this objection,\ but to lay the groundwork for a rigorous discussion, and thus
hopefully lead to new science. \ Section \ref{QMFAIL} provides the
philosophical motivation for our paper, by examining the arguments of several
quantum computing skeptics, including Leonid Levin, Gerard 't Hooft, and
Stephen Wolfram. \ It concludes that a key weakness of their arguments is
their failure to answer the following question: \textit{Exactly what property
separates the quantum states we are sure we can create, from those that
suffice for Shor's factoring algorithm? \ }We call such a property a
\textit{Sure/Shor separator}. \ Section \ref{CQS} develops a complexity theory
of pure quantum states, that studies possible Sure/Shor separators. \ In
particular, it introduces \textit{tree states}, which informally are those
states $\left\vert \psi\right\rangle \in\mathcal{H}_{2}^{\otimes n}%
$\ expressible by a polynomial-size `tree' of addition and tensor product
gates. \ For example, $\alpha\left\vert 0\right\rangle ^{\otimes n}%
+\beta\left\vert 1\right\rangle ^{\otimes n}$\ and $\left(  \alpha\left\vert
0\right\rangle +\beta\left\vert 1\right\rangle \right)  ^{\otimes n}$ are both
tree states. \ Section \ref{BASIC}\ investigates basic properties of this
class of states. \ Among other results, it shows that any tree state is
representable by a tree of polynomial size and logarithmic depth; and that
most states do not even have large inner product with any tree state.

Our main results, proved in Section \ref{LOWER}, are lower bounds on tree size
for various natural families of quantum states. \ In particular, Section
\ref{ECC} analyzes \textquotedblleft subgroup states,\textquotedblright\ which
are uniform superpositions $\left\vert S\right\rangle $\ over all elements of
a subgroup $S\leq\mathbb{Z}_{2}^{n}$. \ The importance of these states arises
from their central role in stabilizer codes, a type of quantum
error-correcting code. \ We first show that if $S$ is chosen uniformly at
random, then with high probability $\left\vert S\right\rangle $ cannot be
represented by any tree of size $n^{o\left(  \log n\right)  }$. \ This result
has a corollary of independent complexity-theoretic interest: the first
superpolynomial gap between the formula size and the multilinear formula size
of a function $f:\left\{  0,1\right\}  ^{n}\rightarrow\mathbb{R}$. \ We then
present two improvements of our basic lower bound. \ First, we show that a
random subgroup state cannot even be \textit{approximated} well in trace
distance by any tree of size $n^{o\left(  \log n\right)  }$. \ Second, we
\textquotedblleft derandomize\textquotedblright\ the lower bound, by using
Reed-Solomon codes to construct an \textit{explicit} subgroup state\ with tree
size $n^{\Omega\left(  \log n\right)  }$.

Section \ref{DIVIS} analyzes the states that arise in Shor's factoring
algorithm---for example, a uniform superposition over all multiples of a fixed
positive integer $p$, written in binary.\ \ Originally, we had hoped to show a
superpolynomial tree size lower bound for these states as well. \ However, we
are only able to show such a bound assuming a number-theoretic conjecture.

Our lower bounds use a sophisticated recent technique of Raz \cite{raz,raz2},
which was introduced to show that the permanent and determinant of a matrix
require superpolynomial-size multilinear formulas. \ Currently, Raz's
technique is only able to show\ lower bounds of the form $n^{\Omega\left(
\log n\right)  }$, but we conjecture that $2^{\Omega\left(  n\right)  }%
$\ lower bounds hold in all of the cases discussed above.

One might wonder how tree size relates to more \textit{physical} properties of
quantum states, such as their robustness to decoherence. \ Section
\ref{PERSIST}\ addresses this question. \ In particular, it shows that if
$\left\vert S\right\rangle $\ is a superposition over codewords of any
sufficiently good erasure code, then $\left\vert S\right\rangle $\ has tree
size $n^{\Omega\left(  \log n\right)  }$, although not vice versa. \ It also
argues that Raz's lower bound technique is connected to a notion called
\textquotedblleft persistence of entanglement,\textquotedblright\ but gives
examples showing that the connection is not exact.

Section \ref{TREEBQP} addresses the following question. \ If the state of a
quantum computer at every time step is a tree state, then can the computer be
simulated classically? \ In other words, letting $\mathsf{TreeBQP}$\ be the
class of languages accepted by such a machine, does $\mathsf{TreeBQP}%
=\mathsf{BPP}$? \ A positive answer would make tree states more attractive as
a Sure/Shor separator. \ For once we admit any states incompatible with the
polynomial-time Church-Turing thesis, it seems like we might as well go all
the way, and admit \textit{all} states preparable by polynomial-size quantum
circuits! \ Although we leave this question open, we do show that
$\mathsf{TreeBQP}\subseteq\mathsf{\Sigma}_{3}^{\mathsf{P}}\cap\mathsf{\Pi}%
_{3}^{\mathsf{P}}$, where $\mathsf{\Sigma}_{3}^{\mathsf{P}}\cap\mathsf{\Pi
}_{3}^{\mathsf{P}}$\ is the third level of the polynomial hierarchy
$\mathsf{PH}$. \ By contrast, it is conjectured that $\mathsf{BQP}%
\not \subset \mathsf{PH}$, though admittedly not on strong evidence.

Section \ref{EXPER}\ discusses the implications of our results for
experimental physics. \ It advocates a dialectic between theory and
experiment, in which theorists would propose a class of quantum states that
encompasses everything seen so far, and then experimenters would try to
prepare states not in that class. \ It also asks whether states with
superpolynomial tree size have already been observed in condensed-matter
systems; and more broadly, what sort of evidence is needed to establish a
state's existence. \ Other issues addressed in Section \ref{EXPER}\ include
how to deal with mixed states and particle position and momentum states, and
the experimental relevance of asymptotic bounds.

Finally, two appendices investigate quantum state complexity measures other
than tree size. \ Appendix \ref{CQS}\ shows relationships among tree size,
circuit size, bounded-depth tree size, Vidal's $\chi$\ complexity
\cite{vidal}, and several other measures. \ It also relates questions about
quantum state classes to more traditional questions about computational
complexity classes. \ Appendix \ref{MOTS}\ studies a weakening of tree size
called \textquotedblleft manifestly orthogonal tree size,\textquotedblright%
\ and shows that this measure can sometimes be characterized \textit{exactly},
enabling us to prove \textit{exponential} lower bounds. \ Our techniques in
Appendix \ref{MOTS}\ might be of independent interest to complexity theorists.

We conclude in Section \ref{OPEN}\ with some open problems.

\section{How Quantum Mechanics Could Fail\label{QMFAIL}}

This section discusses objection (A), that quantum computing is impossible for
a fundamental physical reason. \ Among computer scientists, this objection is
most closely associated with Leonid Levin \cite{levin}.\footnote{Since this
paper was written, Oded Goldreich \cite{goldreich} has also put forward an
argument against quantum computing. Compared to Levin's arguments, Goldreich's
is easily understood: he believes that Shor states have exponential
\textquotedblleft non-degeneracy\textquotedblright\ and therefore take
exponential time to prepare, and that there is no burden on those who hold
this view to suggest a definition of non-degeneracy.} \ The following passage
captures much of the flavor of his critique:

\begin{quote}
The major problem [with quantum computing] is the requirement that basic
quantum equations hold to multi-hundredth if not millionth decimal positions
where the significant digits of the relevant quantum amplitudes reside. \ We
have never seen a physical law valid to over a dozen decimals. \ Typically,
every few new decimal places require major rethinking of most basic concepts.
\ Are quantum amplitudes still complex numbers to such accuracies or do they
become quaternions, colored graphs, or sick-humored gremlins?\ \cite{levin}
\end{quote}

Among other things, Levin argues that quantum computing is analogous to the
unit-cost arithmetic model, and should be rejected for essentially the same
reasons; that claims to the contrary rest on a confusion between metric and
topological approximation; that quantum fault-tolerance theorems depend on
extravagant assumptions; and that even if a quantum computer failed, we could
not measure its state to prove a breakdown of quantum mechanics, and thus
would be unlikely to learn anything new.

A few responses to Levin's arguments can be offered immediately. \ First, even
classically, one can flip a coin a thousand times to produce probabilities of
order $2^{-1000}$. \ Should one dismiss such probabilities as unphysical? \ At
the very least, it is not obvious that amplitudes should behave differently
than probabilities with respect to error---since both evolve linearly, and
neither is directly observable.

Second, if Levin believes that quantum mechanics will fail, but is agnostic
about what will replace it, then his argument can be turned around. \ How do
we know that the successor to quantum mechanics will limit us to
$\mathsf{BPP}$, rather than letting us solve (say) $\mathsf{PSPACE}$-complete
problems? \ This is more than a logical point. \ Abrams and Lloyd
\cite{al}\ argue that a wide class of nonlinear variants of the
Schr\"{o}dinger equation would allow $\mathsf{NP}$-complete and even
$\mathsf{\#P}$-complete problems to be solved\ in polynomial time. \ And
Penrose \cite{penrose}, who proposed a model for `objective collapse' of the
wavefunction, believes that his proposal takes us outside the set of
computable functions entirely!

Third, to falsify quantum mechanics, it would suffice to show that a quantum
computer evolved to \textit{some} state far from the state that quantum
mechanics predicts. \ Measuring the exact state is unnecessary. \ Nobel prizes
have been awarded in the past `merely' for falsifying a previously held
theory, rather than replacing it by a new one. \ An example is the physics
Nobel awarded to Fitch \cite{fitch} and Cronin\ \cite{cronin}\ in 1980 for
discovering CP symmetry violation.

Perhaps the key to understanding Levin's unease about quantum computing lies
in his remark that \textquotedblleft we have never seen a physical law valid
to over a dozen decimals.\textquotedblright\ \ Here he touches on a serious
epistemological question:\ \textit{How far should we extrapolate from today's
experiments to where quantum mechanics has never been tested?} \ We will try
to address this question by reviewing the evidence for quantum mechanics.
\ For our purposes it will not suffice to declare the predictions of quantum
mechanics \textquotedblleft verified to one part in a
trillion,\textquotedblright\ because we need to distinguish at least three
different \textit{types} of prediction: \textit{interference},
\textit{entanglement}, and \textit{Schr\"{o}dinger cats}. \ Let us consider
these in turn.

\begin{enumerate}
\item[(1)] \textbf{Interference.} \ If the different paths that an electron
could take in its orbit around a nucleus did not interfere destructively,
canceling each other out, then electrons would not have quantized energy
levels. \ So being accelerating electric charges, they would lose energy and
spiral into their respective nuclei, and all matter would disintegrate. \ That
this has not happened---together with the results of (for example)
single-photon double-slit experiments---is compelling evidence for the reality
of quantum interference.

\item[(2)] \textbf{Entanglement.} \ One might accept that a single particle's
position is described by a wave in three-dimensional phase space, but deny
that two particles are described by a wave in \textit{six}-dimensional phase
space.\ \ However, the Bell inequality experiments of Aspect et al.
\cite{aspect}\ and successors have convinced all but a few physicists that
quantum entanglement exists, can be maintained over large distances, and
cannot be explained by local hidden-variable theories.

\item[(3)] \textbf{Schr\"{o}dinger Cats.} \ Accepting two- and three-particle
entanglement is not the same as accepting that whole molecules, cats, humans,
and galaxies can be in coherent superposition states. \ However, recently
Arndt et al. \cite{arndt}\ have performed the double-slit interference
experiment using $C_{60}$\ molecules (buckyballs) instead of photons; while
Friedman et al. \cite{friedman}\ have found evidence that a superconducting
current, consisting of billions of electrons, can enter a coherent
superposition of flowing clockwise around a coil and flowing counterclockwise
(see Leggett \cite{leggett} for a survey of such experiments). \ Though short
of cats, these experiments at least allow us to say the following: \textit{if
we could build a general-purpose quantum computer with as many components as
have already been placed into coherent superposition, then on certain
problems, that computer would outperform any computer in the world today.}
\end{enumerate}

Having reviewed some of the evidence for quantum mechanics, we must now ask
what alternatives have been proposed that might also explain the evidence.
\ The simplest alternatives are those in which quantum states
\textquotedblleft spontaneously collapse\textquotedblright\ with some
probability, as in the GRW (Ghirardi-Rimini-Weber) theory \cite{grw}%
.\footnote{Penrose \cite{penrose}\ has proposed another such theory, but as
mentioned earlier, his theory suggests that the quantum computing model is
\textit{too} restrictive.}\ \ The drawbacks of the GRW theory include
violations of energy conservation, and parameters that must be fine-tuned to
avoid conflicting with experiments. \ More relevant for us, though, is that
the collapses postulated by the theory are only in the position basis, so that
quantum information stored in internal degrees of freedom (such as spin) is
unaffected. \ Furthermore, even if we extended the theory to collapse those
internal degrees, large quantum computers could still be built. \ For the
theory predicts roughly one collapse per particle per $10^{15}$\ seconds, with
a collapse affecting everything in a $10^{-7}$-meter vicinity. \ So even in
such a vicinity, one could perform a computation involving (say) $10^{10}%
$\ particles for $10^{5}$\ seconds. \ Finally, as pointed out to us by Rob
Spekkens, standard quantum error-correction techniques might be used to
overcome even GRW-type decoherence.

A second class of alternatives includes those of 't Hooft \cite{thooft}\ and
Wolfram \cite{wolfram}, in which something like a deterministic cellular
automaton underlies quantum mechanics. \ On the basis of his theory, 't Hooft
predicts that \textquotedblleft\lbrack i]t will never be possible to construct
a `quantum computer' that can factor a large number faster, and within a
smaller region of space, than a classical machine would do, if the latter
could be built out of parts at least as large and as slow as the Planckian
dimensions\textquotedblright\ \cite{thooft}. \ Similarly, Wolfram states\ that
``[i]ndeed within the usual formalism [of quantum mechanics] one can construct
quantum computers that may be able to solve at least a few specific problems
exponentially faster than ordinary Turing machines. \ But particularly after
my discoveries \ldots\ I strongly suspect that even if this is formally the
case, it will still not turn out to be a true representation of ultimate
physical reality, but will instead just be found to reflect various
idealizations made in the models used so far'' \cite[p.771]{wolfram}.

The obvious question then is how these theories account for Bell inequality
violations. \ We confess to being unable to understand 't Hooft's answer to
this question, except that he believes that the usual notions of causality and
locality might no longer apply in quantum gravity. \ As for Wolfram's theory,
which involves \textquotedblleft long-range threads\textquotedblright\ to
account for Bell inequality violations, we argued in \cite{aarrev}\ that it
fails Wolfram's own desiderata of causal and relativistic invariance.

So the challenge for quantum computing skeptics is clear. \ Ideally, come up
with an alternative to quantum mechanics---even an idealized toy theory---that
can account for all present-day experiments, yet would not allow large-scale
quantum computation. \ Failing that, \textit{at least say what you take
quantum mechanics' domain of validity to be}. \ One way to do this would be to
propose a set $S$ of quantum states that you believe corresponds to possible
physical states of affairs.\footnote{A skeptic might also specify what happens
if a state $\left\vert \psi\right\rangle \in S$\ is acted on by a unitary $U$
such that $U\left\vert \psi\right\rangle \notin S$, but this will not be
insisted upon.} \ The set $S$ must contain all \textquotedblleft Sure
states\textquotedblright\ (informally, the states that have already been
demonstrated in the lab), but no \textquotedblleft Shor
states\textquotedblright\ (again informally, the states that can be shown to
suffice for factoring, say, $500$-digit numbers). \ If $S$ satisfies both of
these constraints, then we call $S$ a \textit{Sure/Shor separator} (see Figure
1).\begin{figure}[ptb]
\begin{center}
\includegraphics[
trim=0.9in 2.393715in 0.9in 2.5in,
height=2.4in,
width=2.6in
]{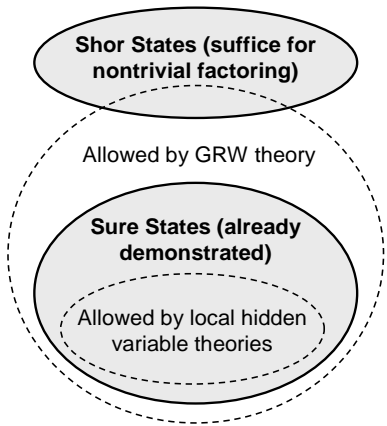}
\end{center}
\caption{A Sure/Shor separator must contain all Sure states but no Shor
states. \ That is why neither local hidden variables nor the GRW theory yields
a Sure/Shor separator.}%
\label{sureshor}%
\end{figure}

Of course, an alternative theory need not involve a sharp cutoff between
possible and impossible states. \ So it is perfectly acceptable for a skeptic
to define a \textquotedblleft complexity measure\textquotedblright\ $C\left(
\left\vert \psi\right\rangle \right)  $\ for quantum states, and then say
something like the following: \ \textit{If }$\left\vert \psi_{n}\right\rangle
$\textit{\ is a state of }$n$\textit{ spins, and }$C\left(  \left\vert
\psi_{n}\right\rangle \right)  $\textit{\ is at most, say, }$n^{2}$\textit{,
then I predict that }$\left\vert \psi_{n}\right\rangle $\textit{\ can be
prepared using only \textquotedblleft polynomial effort.\textquotedblright%
\ \ Also, once prepared, }$\left\vert \psi_{n}\right\rangle $\textit{\ will be
governed by standard quantum mechanics to extremely high precision. \ All
states created to date have had small values of }$C\left(  \left\vert \psi
_{n}\right\rangle \right)  $\textit{. \ However, if }$C\left(  \left\vert
\psi_{n}\right\rangle \right)  $\textit{\ grows as, say, }$2^{n}$\textit{,
then I predict that }$\left\vert \psi_{n}\right\rangle $\textit{\ requires
\textquotedblleft exponential effort\textquotedblright\ to prepare, or else is
not even approximately governed by quantum mechanics, or else does not even
make sense in the context of an alternative theory. \ The states that arise in
Shor's factoring algorithm have exponential values of }$C\left(  \left\vert
\psi_{n}\right\rangle \right)  $\textit{. \ So as my Sure/Shor separator, I
propose the set of all infinite families of states }$\left\{  \left\vert
\psi_{n}\right\rangle \right\}  _{n\geq1}$\textit{, where }$\left\vert
\psi_{n}\right\rangle $\textit{\ has }$n$\textit{ qubits, such that }$C\left(
\left\vert \psi_{n}\right\rangle \right)  \leq p\left(  n\right)  $\textit{
for some polynomial }$p$\textit{.}

To understand the importance of Sure/Shor separators, it is helpful to think
through some examples. \ A major theme of Levin's arguments was that
exponentially small amplitudes are somehow unphysical. \ However, clearly we
cannot reject \textit{all} states with tiny amplitudes---for would anyone
dispute that the state $2^{-5000}\left(  \left\vert 0\right\rangle +\left\vert
1\right\rangle \right)  ^{\otimes10000}$\ is formed whenever $10,000$ photons
are each polarized at $45^{\circ}$?\ \ Indeed, once we accept $\left\vert
\psi\right\rangle $\ and $\left\vert \varphi\right\rangle $\ as Sure states,
we are almost \textit{forced} to accept $\left\vert \psi\right\rangle
\otimes\left\vert \varphi\right\rangle $\ as well---since we can imagine, if
we like, that $\left\vert \psi\right\rangle $\ and $\left\vert \varphi
\right\rangle $\ are prepared in two separate laboratories.\footnote{A
reviewer comments that in Chern-Simons theory (for example), there is no clear
tensor product decomposition. \ However, the only question that concerns us is
whether $\left\vert \psi\right\rangle \otimes\left\vert \varphi\right\rangle
$\ is a Sure state, \textit{given} that $\left\vert \psi\right\rangle $\ and
$\left\vert \varphi\right\rangle $\ are both Sure states that are
well-described in tensor product Hilbert spaces.} \ So considering a Shor
state such as%
\[
\left\vert \Phi\right\rangle =\frac{1}{2^{n/2}}\sum_{r=0}^{2^{n}-1}\left\vert
r\right\rangle \left\vert x^{r}\operatorname{mod}N\right\rangle ,
\]
what property of this state could quantum computing skeptics latch onto as
being physically extravagant? \ They might complain that $\left\vert
\Phi\right\rangle $\ involves entanglement across hundreds or thousands of
particles; but as mentioned earlier, there are other states with that same
property, namely the \textquotedblleft Schr\"{o}dinger cats\textquotedblright%
\ $\left(  \left\vert 0\right\rangle ^{\otimes n}+\left\vert 1\right\rangle
^{\otimes n}\right)  /\sqrt{2}$, that should be regarded as Sure states.
\ Alternatively, the skeptics might object to the \textit{combination} of
exponentially small amplitudes with entanglement across hundreds of particles.
\ However, simply viewing a Schr\"{o}dinger cat state in the Hadamard basis
produces an equal superposition over all strings of even parity, which has
both properties. \ We seem to be on a slippery slope leading to all of quantum
mechanics! \ Is there any defensible place to draw a line?

The dilemma above is what led us to propose \textit{tree states} as a possible
Sure/Shor separator. \ The idea, which might seem more natural to logicians
than to physicists, is this. \ Once we accept the linear combination and
tensor product rules of quantum mechanics---allowing $\alpha\left\vert
\psi\right\rangle +\beta\left\vert \varphi\right\rangle $\ and $\left\vert
\psi\right\rangle \otimes\left\vert \varphi\right\rangle $\ into our set $S$
of possible states whenever $\left\vert \psi\right\rangle ,\left\vert
\varphi\right\rangle \in S$---one of our few remaining hopes for keeping $S$ a
proper subset of the set of \textit{all} states is to impose some restriction
on how those two rules can be iteratively applied. \ In particular, we could
let $S$ be the closure of $\left\{  \left\vert 0\right\rangle ,\left\vert
1\right\rangle \right\}  $\ under a \textit{polynomial number} of linear
combinations and tensor products. \ That is, $S$\ is the set of all infinite
families of states $\left\{  \left\vert \psi_{n}\right\rangle \right\}
_{n\geq1}$ with $\left\vert \psi_{n}\right\rangle \in\mathcal{H}_{2}^{\otimes
n}$, such that $\left\vert \psi_{n}\right\rangle $\ can be expressed as a
\textquotedblleft tree\textquotedblright\ involving at most $p\left(
n\right)  $\ addition, tensor product, $\left\vert 0\right\rangle $, and
$\left\vert 1\right\rangle $\ gates for some polynomial $p$ (see Figure
2).\begin{figure}[ptb]
\begin{center}
\includegraphics[
trim=2in 4.5in 2in -0.35in,
height=1.7in,
width=2.9464in
]{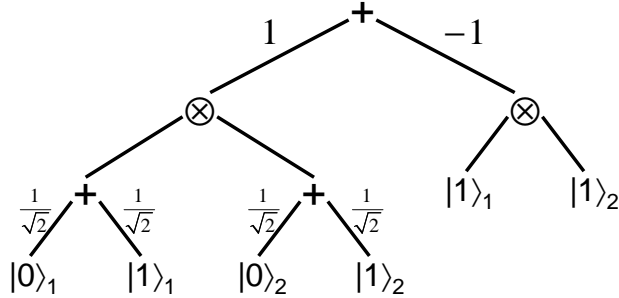}
\end{center}
\caption{Expressing $\left(  \left\vert 00\right\rangle +\left\vert
01\right\rangle +\left\vert 10\right\rangle -\left\vert 11\right\rangle
\right)  /2$\ by a tree of linear combination and tensor product gates, with
scalar multiplication along edges. \ Subscripts denote the identity of a
qubit.}%
\label{tree}%
\end{figure}

To be clear, we are \textit{not} advocating that \textquotedblleft all states
in Nature are tree states\textquotedblright\ as a serious physical
hypothesis.\ \ Indeed, even if we believed firmly in a breakdown of quantum
mechanics,\footnote{which we don't} there are other choices for the set $S$
that seem equally reasonable. \ For example, define \textit{orthogonal tree
states} similarly to tree states, except that we can only form the linear
combination $\alpha\left\vert \psi\right\rangle +\beta\left\vert
\varphi\right\rangle $\ if $\left\langle \psi|\varphi\right\rangle =0$.
\ Rather than choose among tree states, orthogonal tree states, and the other
candidate Sure/Shor separators that occurred to us, our approach will be to
prove everything we can about all of them. \ If we devote more space to tree
states than to others, that is simply because tree states are the subject of
our most interesting results. \ On the other hand, if we show (for example)
that $\left\{  \left\vert \psi_{n}\right\rangle \right\}  $\ is not a tree
state, then we have also shown that $\left\{  \left\vert \psi_{n}\right\rangle
\right\}  $\ is not an orthogonal tree state. \ So many candidate separators
are related to each other; and indeed, their relationships will be a major
theme of the paper.

Let us summarize. \ To debate whether quantum computing is fundamentally
impossible, we need at least one proposal for how it \textit{could} be
impossible. \ Since even skeptics admit that quantum mechanics is valid within
some \textquotedblleft regime,\textquotedblright\ a key challenge for any such
proposal is to separate the regime of acknowledged validity from the quantum
computing regime. \ Though others will disagree, we do not see any choice but
to \textit{identify those two regimes with classes of quantum states}. \ For
gates and measurements that suffice for quantum computing have already been
demonstrated experimentally. \ Thus, if we tried to identify the two regimes
with classes of gates or measurements, then we could equally well talk about
the class of \textit{states} on which all $1$- and $2$-qubit operations behave
as expected. \ A similar argument would apply if we identified the two regimes
with classes of quantum circuits---since any \textquotedblleft
memory\textquotedblright\ that a quantum system retains of the previous gates
in a circuit, is part of the system's state by definition. \ So: states,
gates, measurements, circuits---what else is there?

We should stress that none of the above depends on the interpretation of
quantum mechanics. \ In particular, it is irrelevant whether we regard quantum
states as \textquotedblleft really out there\textquotedblright\ or as
representing subjective knowledge---since in either case, the question is
whether there can exist systems that we would \textit{describe} by $\left\vert
\psi\right\rangle $\ based on their observed behavior.

Once we agree to seek a Sure/Shor separator, we quickly find that the obvious
ideas---based on precision in amplitudes, or entanglement across of hundreds
of particles---are nonstarters. \ The only idea that we have found plausible
is to limit the class of allowed quantum states to those with some kind of
succinct representation. \ That still leaves numerous possibilities; and for
each one, it might be a difficult problem to decide whether a given
$\left\vert \psi\right\rangle $\ is succinctly representable or not. \ Thus,
constructing a useful theory of Sure/Shor separators will not be easy. \ But
we should start somewhere.

\section{Classifying Quantum States\label{CQS}}

In both quantum and classical complexity theory, the objects studied are
usually sets of languages or Boolean functions. \ However, a generic $n$-qubit
quantum state requires exponentially many classical bits to describe, and this
suggests looking at \textit{the complexity of quantum states themselves}.
\ That is, which states have polynomial-size classical descriptions of various
kinds? \ This question has been studied from several angles by Aharonov and
Ta-Shma \cite{at};\ Janzing, Wocjan, and Beth \cite{jwb};\ Vidal \cite{vidal};
and Green et al. \cite{ghmp}. \ Here we propose a general framework for the
question. \ For simplicity, we limit ourselves to pure states $\left\vert
\psi_{n}\right\rangle \in\mathcal{H}_{2}^{\otimes n}$ with the fixed
orthogonal basis $\left\{  \left\vert x\right\rangle :x\in\left\{
0,1\right\}  ^{n}\right\}  $. \ Also, by `states' we mean infinite families of
states $\left\{  \left\vert \psi_{n}\right\rangle \right\}  _{n\geq1}$.

Like complexity classes, pure quantum states can be organized into a hierarchy
(see Figure 3). \ At the bottom are the classical basis states, which have the
form $\left\vert x\right\rangle $\ for some $x\in\left\{  0,1\right\}  ^{n}$.
\ We can generalize classical states in two directions: to the class
$\mathsf{\otimes}_{\mathsf{1}}$ of separable states, which have the form
$\left(  \alpha_{1}\left\vert 0\right\rangle +\beta_{1}\left\vert
1\right\rangle \right)  \otimes\cdots\otimes\left(  \alpha_{n}\left\vert
0\right\rangle +\beta_{n}\left\vert 1\right\rangle \right)  $; and to the
class $\mathsf{\Sigma}_{\mathsf{1}}$, which consists of all states $\left\vert
\psi_{n}\right\rangle $\ that are superpositions of at most $p\left(
n\right)  $\ classical states, where $p$ is a polynomial. \ At the next level,
$\mathsf{\otimes}_{\mathsf{2}}$ contains the states that can be written as a
tensor product of $\mathsf{\Sigma}_{\mathsf{1}}$\ states, with qubits permuted
arbitrarily. \ Likewise, $\mathsf{\Sigma}${}$_{\mathsf{2}}$ contains the
states that can be written as a linear combination of a polynomial number of
$\mathsf{\otimes}_{\mathsf{1}}$ states. \ We can continue indefinitely to
$\mathsf{\Sigma}${}$_{\mathsf{3}}$, $\mathsf{\otimes}_{\mathsf{3}}$, etc.
\ Containing the whole `tensor-sum hierarchy' $\mathsf{\cup}_{\mathsf{k}%
}\mathsf{\Sigma}${}$_{\mathsf{k}}=\mathsf{\cup}_{\mathsf{k}}\mathsf{\otimes
}_{\mathsf{k}}$\ is the class $\mathsf{Tree}$, of all states expressible by a
polynomial-size tree of additions and tensor products nested arbitrarily.
\ Formally, $\mathsf{Tree}$\ consists of all states $\left\vert \psi
_{n}\right\rangle $\ such that $\operatorname*{TS}\left(  \left\vert \psi
_{n}\right\rangle \right)  \leq p\left(  n\right)  $\ for some polynomial
$p$,\ where the \textit{tree size} $\operatorname*{TS}\left(  \left\vert
\psi_{n}\right\rangle \right)  $\ is defined as follows.\begin{figure}[ptb]
\begin{center}
\includegraphics[
trim=0.678569in 1.806069in 0.679377in 1.813617in,
height=2.8928in,
width=2.7164in
]{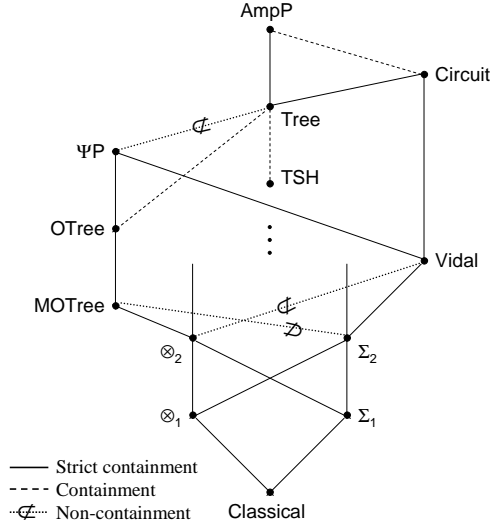}
\end{center}
\caption{Relations among quantum state classes.}%
\label{classes}%
\end{figure}

\begin{definition}
A \textit{quantum state tree} over $\mathcal{H}_{2}^{\otimes n}$\ is a rooted
tree where each leaf vertex is labeled with $\alpha\left|  0\right\rangle
+\beta\left|  1\right\rangle $ for some $\alpha,\beta\in\mathsf{C}$, and each
non-leaf vertex (called a gate) is labeled with either $+$ or $\otimes$.
\ Each vertex $v$ is also labeled with a set $S\left(  v\right)
\subseteq\left\{  1,\ldots,n\right\}  $, such that

\begin{enumerate}
\item[(i)] If $v$ is a leaf then $\left|  S\left(  v\right)  \right|  =1$,

\item[(ii)] If $v$ is the root then $S\left(  v\right)  =\left\{
1,\ldots,n\right\}  $,

\item[(iii)] If $v$ is a $+$ gate and $w$ is a child of $v$, then $S\left(
w\right)  =S\left(  v\right)  $,

\item[(iv)] If $v$ is a $\otimes$\ gate and $w_{1},\ldots,w_{k}$\ are the
children of $v$, then $S\left(  w_{1}\right)  ,\ldots,S\left(  w_{k}\right)
$\ are pairwise disjoint and form a partition of $S\left(  v\right)  $.
\end{enumerate}

Finally, if $v$\ is a $+$ gate, then the outgoing edges\ of $v$ are labeled
with complex numbers. \ For each $v$, the subtree rooted at $v$ represents a
quantum state of the qubits in $S\left(  v\right)  $\ in the obvious way. \ We
require this state to be normalized for each $v$.\footnote{Requiring only the
\textit{whole} tree to represent a normalized state clearly yields no further
generality.}
\end{definition}

We say a tree is \textit{orthogonal} if it satisfies the further condition
that if $v$ is a $+$ gate, then any two children $w_{1},w_{2}$\ of $v$
represent $\left\vert \psi_{1}\right\rangle ,\left\vert \psi_{2}\right\rangle
$\ with $\left\langle \psi_{1}|\psi_{2}\right\rangle =0$. \ If the condition
$\left\langle \psi_{1}|\psi_{2}\right\rangle =0$ can be replaced by the
stronger condition that for all basis states $\left\vert x\right\rangle $,
either $\left\langle \psi_{1}|x\right\rangle =0$ or $\left\langle \psi
_{2}|x\right\rangle =0$, then we say the tree is \textit{manifestly
orthogonal}. \ Manifest orthogonality is an extremely unphysical definition;
we introduce it only because it is interesting from a lower bounds perspective.

For reasons of convenience, we define the \textit{size} $\left\vert
T\right\vert $\ of a tree $T$ to be the number of leaf vertices. \ Then given
a state $\left\vert \psi\right\rangle \in\mathcal{H}_{2}^{\otimes n}$, the
\textit{tree size} $\operatorname*{TS}\left(  \left\vert \psi\right\rangle
\right)  $\ is the minimum size of a tree that represents $\left\vert
\psi\right\rangle $. \ The \textit{orthogonal tree size} $\operatorname*{OTS}%
\left(  \left\vert \psi\right\rangle \right)  $\ and \textit{manifestly
orthogonal tree size} $\operatorname*{MOTS}\left(  \left\vert \psi
\right\rangle \right)  $\ are defined similarly. \ Then $\mathsf{OTree}$\ is
the class of $\left\vert \psi_{n}\right\rangle $\ such that
$\operatorname*{OTS}\left(  \left\vert \psi_{n}\right\rangle \right)  \leq
p\left(  n\right)  $ for some polynomial $p$, and $\mathsf{MOTree}$\ is the
class such that $\operatorname*{MOTS}\left(  \left\vert \psi_{n}\right\rangle
\right)  \leq p\left(  n\right)  $\ for some $p$.

It is easy to see that
\[
n\leq\operatorname*{TS}\left(  \left\vert \psi\right\rangle \right)
\leq\operatorname*{OTS}\left(  \left\vert \psi\right\rangle \right)
\leq\operatorname*{MOTS}\left(  \left\vert \psi\right\rangle \right)  \leq
n2^{n}%
\]
for every $\left\vert \psi\right\rangle $, and that the set of $\left\vert
\psi\right\rangle $\ such that $\operatorname*{TS}\left(  \left\vert
\psi\right\rangle \right)  <2^{n}$\ has measure $0$ in $\mathcal{H}%
_{2}^{\otimes n}$. \ Two other important properties of $\operatorname*{TS}$
and $\operatorname*{OTS}$ are as follows:

\begin{proposition}
\label{invariant}\quad

\begin{enumerate}
\item[(i)] $\operatorname*{TS}$ and $\operatorname*{OTS}$ are invariant under
local\footnote{Several people told us that a reasonable complexity measure
must be invariant under \textit{all} basis changes. Alas, this would imply
that all pure states have the same complexity!} basis changes, up to a
constant factor of $2$.

\item[(ii)] If $\left\vert \phi\right\rangle $ is obtained from $\left\vert
\psi\right\rangle $ by applying a $k$-qubit unitary, then $\operatorname*{TS}%
\left(  \left\vert \phi\right\rangle \right)  \leq k4^{k}\operatorname*{TS}%
\left(  \left\vert \psi\right\rangle \right)  $ and $\operatorname*{OTS}%
\left(  \left\vert \phi\right\rangle \right)  \leq k4^{k}\operatorname*{OTS}%
\left(  \left\vert \psi\right\rangle \right)  $.
\end{enumerate}
\end{proposition}

\begin{proof}
\quad

\begin{enumerate}
\item[(i)] Simply replace each occurrence of $\left\vert 0\right\rangle $\ in
the original tree by a tree for $\alpha\left\vert 0\right\rangle
+\beta\left\vert 1\right\rangle $, and each occurrence of $\left\vert
1\right\rangle $\ by a tree for $\gamma\left\vert 0\right\rangle
+\delta\left\vert 1\right\rangle $, as appropriate.

\item[(ii)] Suppose without loss of generality that the gate is applied to the
first $k$ qubits. \ Let $T$ be a tree representing $\left\vert \psi
\right\rangle $, and let $T_{y}$\ be the restriction of $T$\ obtained by
setting the first $k$ qubits to $y\in\left\{  0,1\right\}  ^{k}$. \ Clearly
$\left\vert T_{y}\right\vert \leq\left\vert T\right\vert $. \ Furthermore, we
can express $\left\vert \phi\right\rangle $\ in the form $\sum_{y\in\left\{
0,1\right\}  ^{k}}S_{y}T_{y}$, where each $S_{y}$\ represents a $k$-qubit
state and hence is expressible by a tree of size $k2^{k}$.
\end{enumerate}
\end{proof}

We can also define the $\varepsilon$\textit{-approximate tree size}
$\operatorname*{TS}_{\varepsilon}\left(  \left|  \psi\right\rangle \right)
$\ to be the minimum size of a tree representing a state $\left|
\varphi\right\rangle $\ such that $\left|  \left\langle \psi|\varphi
\right\rangle \right|  ^{2}\geq1-\varepsilon$, and define $\operatorname*{OTS}%
_{\varepsilon}\left(  \left|  \psi\right\rangle \right)  $\ and
$\operatorname*{MOTS}_{\varepsilon}\left(  \left|  \psi\right\rangle \right)
$\ similarly.

\begin{definition}
An arithmetic formula (over the ring $\mathbb{C}$ and $n$ variables) is a
rooted binary tree where each leaf vertex is labeled with either a complex
number or a variable in $\left\{  x_{1},\ldots,x_{n}\right\}  $, and each
non-leaf vertex is labeled with either $+$\ or $\times$. \ Such a tree
represents a polynomial $p\left(  x_{1},\ldots,x_{n}\right)  $\ in the obvious
way. \ We call a polynomial \textit{multilinear} if no variable appears raised
to a higher power than $1$, and an arithmetic formula multilinear if the
polynomials computed by each of its subtrees are multilinear.
\end{definition}

The \textit{size} $\left\vert \Phi\right\vert $\ of a multilinear formula
$\Phi$\ is the number of leaf vertices. \ Given a multilinear polynomial $p$,
the multilinear formula size $\operatorname*{MFS}\left(  p\right)  $\ is the
minimum size of a multilinear formula that represents $p$. \ Then given a
function $f:\left\{  0,1\right\}  ^{n}\rightarrow\mathbb{C}$, we define
\[
\operatorname*{MFS}\left(  f\right)  =\min_{p~:~p\left(  x\right)  =f\left(
x\right)  ~\forall x\in\left\{  0,1\right\}  ^{n}}\operatorname*{MFS}\left(
p\right)  .
\]
(Actually $p$ turns out to be unique \cite{ns}.) \ We can also define the
$\varepsilon$-approximate\ multilinear formula size of $f$,
\[
\operatorname*{MFS}\nolimits_{\varepsilon}\left(  f\right)  =\min
_{p~:~\left\Vert p-f\right\Vert _{2}^{2}\leq\varepsilon}\operatorname*{MFS}%
\left(  p\right)
\]
where $\left\Vert p-f\right\Vert _{2}^{2}=\sum_{x\in\left\{  0,1\right\}
^{n}}\left\vert p\left(  x\right)  -f\left(  x\right)  \right\vert ^{2}$.
\ (This metric is closely related to the inner product $\sum_{x}p\left(
x\right)  ^{\ast}f\left(  x\right)  $, but is often more convenient to work
with.) \ Now given a state $\left\vert \psi\right\rangle =\sum_{x\in\left\{
0,1\right\}  ^{n}}\alpha_{x}\left\vert x\right\rangle $ in $\mathcal{H}%
_{2}^{\otimes n}$, let $f_{\psi}$\ be the function from $\left\{  0,1\right\}
^{n}$\ to $\mathbb{C}$\ defined by $f_{\psi}\left(  x\right)  =\alpha_{x}$.

\begin{theorem}
\label{iff}For all $\left\vert \psi\right\rangle $,

\begin{enumerate}
\item[(i)] $\operatorname*{MFS}\left(  f_{\psi}\right)  =O\left(
\operatorname*{TS}\left(  \left\vert \psi\right\rangle \right)  \right)  $.

\item[(ii)] $\operatorname*{TS}\left(  \left|  \psi\right\rangle \right)
=O\left(  \operatorname*{MFS}\left(  f_{\psi}\right)  +n\right)  $.

\item[(iii)] $\operatorname*{MFS}_{\delta}\left(  f_{\psi}\right)  =O\left(
\operatorname*{TS}_{\varepsilon}\left(  \left\vert \psi\right\rangle \right)
\right)  $ where $\delta=2-2\sqrt{1-\varepsilon}$.

\item[(iv)] $\operatorname*{TS}_{2\varepsilon}\left(  \left\vert
\psi\right\rangle \right)  =O\left(  \operatorname*{MFS}_{\varepsilon}\left(
f_{\psi}\right)  +n\right)  $.
\end{enumerate}
\end{theorem}

\begin{proof}

\begin{enumerate}
\item[(i)] Given a tree representing $\left\vert \psi\right\rangle $, replace
every unbounded fan-in gate by a collection of binary gates, every $\otimes
$\ by $\times$, every $\left\vert 1\right\rangle _{i}$\ vertex by $x_{i}$, and
every $\left\vert 0\right\rangle _{i}$\ vertex by a formula for $1-x_{i}$.
\ Push all multiplications by constants at the edges down to $\times$\ gates
at the leaves.

\item[(ii)] Given a multilinear formula $\Phi$ for $f_{\psi}$, let $p\left(
v\right)  $\ be the polynomial computed at vertex $v$ of $\Phi$, and let
$S\left(  v\right)  $\ be the set of variables that appears in $p\left(
v\right)  $. \ First, call $\Phi$ \textit{syntactic} if at every $\times$ gate
with children $v$ and $w$, $S\left(  v\right)  \cap S\left(  w\right)
=\varnothing$. \ A lemma of Raz \cite{raz}\ states that we can always make
$\Phi$ syntactic without increasing its size.

Second, at every $+$\ gate $u$ with children $v$ and $w$, enlarge both
$S\left(  v\right)  $\ and $S\left(  w\right)  $\ to $S\left(  v\right)  \cup
S\left(  w\right)  $, by multiplying $p\left(  v\right)  $\ by $x_{i}+\left(
1-x_{i}\right)  $\ for every $x_{i}\in S\left(  w\right)  \setminus S\left(
v\right)  $, and multiplying $p\left(  w\right)  $\ by $x_{i}+\left(
1-x_{i}\right)  $\ for every $x_{i}\in S\left(  v\right)  \setminus S\left(
w\right)  $. \ Doing this does not invalidate any $\times$\ gate that is an
ancestor of $u$, since by the assumption that $\Phi$ is syntactic, $p\left(
u\right)  $\ is never multiplied by any polynomial containing variables in
$S\left(  v\right)  \cup S\left(  w\right)  $. \ Similarly, enlarge $S\left(
r\right)  $\ to $\left\{  x_{1},\ldots,x_{n}\right\}  $\ where $r$ is the root
of $\Phi$.

Third, call $v$ \textit{max-linear} if $\left\vert S\left(  v\right)
\right\vert =1$\ but $\left\vert S\left(  w\right)  \right\vert >1$ where $w$
is the parent of $v$. \ If $v$ is max-linear and $p\left(  v\right)
=a+bx_{i}$, then replace the tree rooted at $v$ by a tree computing
$a\left\vert 0\right\rangle _{i}+\left(  a+b\right)  \left\vert 1\right\rangle
_{i}$. \ Also, replace all multiplications by constants higher in $\Phi$ by
multiplications at the edges. \ (Because of the second step, there are no
additions by constants higher in $\Phi$.) \ Replacing every $\times$\ by
$\otimes$\ then gives a tree representing $\left\vert \psi\right\rangle $,
whose size is easily seen to be $O\left(  \left\vert \Phi\right\vert
+n\right)  $\ .

\item[(iii)] Apply the reduction from part (i). \ Let the resulting
multilinear formula compute polynomial $p$; then%
\[
\sum_{x\in\left\{  0,1\right\}  ^{n}}\left\vert p\left(  x\right)  -f_{\psi
}\left(  x\right)  \right\vert ^{2}=2-2\sum_{x\in\left\{  0,1\right\}  ^{n}%
}p\left(  x\right)  \overline{f_{\psi}\left(  x\right)  }\leq2-2\sqrt
{1-\varepsilon}=\delta.
\]

\item[(iv)] Apply the reduction from part (ii). \ Let $\left(  \beta
_{x}\right)  _{x\in\left\{  0,1\right\}  ^{n}}$\ be the resulting amplitude
vector; since this vector might not be normalized, divide each $\beta_{x}$ by
$\sum_{x}\left\vert \beta_{x}\right\vert ^{2}$ to produce $\beta_{x}^{\prime}%
$. \ Then%
\begin{align*}
\left\vert \sum_{x\in\left\{  0,1\right\}  ^{n}}\beta_{x}^{\prime}%
\overline{\alpha_{x}}\right\vert ^{2}  &  =1-\frac{1}{2}\sum_{x\in\left\{
0,1\right\}  ^{n}}\left\vert \beta_{x}^{\prime}-\alpha_{x}\right\vert ^{2}\\
&  \geq1-\frac{1}{2}\left(  \sqrt{\sum_{x\in\left\{  0,1\right\}  ^{n}%
}\left\vert \beta_{x}^{\prime}-\beta_{x}\right\vert ^{2}}+\sqrt{\sum
_{x\in\left\{  0,1\right\}  ^{n}}\left\vert \beta_{x}-\alpha_{x}\right\vert
^{2}}\right)  ^{2}\\
&  \geq1-\frac{1}{2}\left(  2\sqrt{\varepsilon}\right)  ^{2}=1-2\varepsilon.
\end{align*}

\end{enumerate}
\end{proof}

Besides $\mathsf{Tree}$, $\mathsf{OTree}$, and $\mathsf{MOTree}$, four other
classes of quantum states deserve mention:

$\mathsf{Circuit}$, a circuit analog of $\mathsf{Tree}$, contains the states
$\left\vert \psi_{n}\right\rangle =\sum_{x}\alpha_{x}\left\vert x\right\rangle
$\ such that for all $n$, there exists a multilinear arithmetic circuit of
size $p\left(  n\right)  $\ over the complex numbers that outputs $\alpha_{x}%
$\ given $x$ as input, for some polynomial $p$. \ (Multilinear circuits are
the same as multilinear trees, except that they allow unbounded fanout---that
is, polynomials computed at intermediate points can be reused arbitrarily many times.)

$\mathsf{AmpP}$ contains the states $\left|  \psi_{n}\right\rangle =\sum
_{x}\alpha_{x}\left|  x\right\rangle $\ such that for all $n,b$, there exists
a classical circuit of size $p\left(  n+b\right)  $\ that outputs $\alpha_{x}$
to $b$ bits of precision given $x$ as input, for some polynomial $p$.

$\mathsf{Vidal}$ contains the states that are `polynomially entangled' in the
sense of Vidal \cite{vidal}. \ Given a partition of $\left\{  1,\ldots
,n\right\}  $\ into $A$ and $B$, let $\chi_{A}\left(  \left\vert \psi
_{n}\right\rangle \right)  $\ be the minimum $k$ for which $\left\vert
\psi_{n}\right\rangle $\ can be written as $\sum_{i=1}^{k}\alpha_{i}\left\vert
\varphi_{i}^{A}\right\rangle \otimes\left\vert \varphi_{i}^{B}\right\rangle $,
where $\left\vert \varphi_{i}^{A}\right\rangle $\ and $\left\vert \varphi
_{i}^{B}\right\rangle $\ are states of qubits in $A$ and $B$
respectively.\ \ \ ($\chi_{A}\left(  \left\vert \psi_{n}\right\rangle \right)
$\ is known as the \textit{Schmidt rank}; see \cite{nc}\ for more
information.) \ Let $\chi\left(  \left\vert \psi_{n}\right\rangle \right)
=\max_{A}\chi_{A}\left(  \left\vert \psi_{n}\right\rangle \right)  $. \ Then
$\left\vert \psi_{n}\right\rangle \in\mathsf{Vidal}$\ if and only if
$\chi\left(  \left\vert \psi_{n}\right\rangle \right)  \leq p\left(  n\right)
$\ for some polynomial $p$.

$\mathsf{\Psi P}$ contains the states $\left\vert \psi_{n}\right\rangle $ such
that for all $n$ and $\varepsilon>0$, there exists a quantum circuit of size
$p\left(  n+\log\left(  1/\varepsilon\right)  \right)  $\ that maps the
all-$0$ state to a state some part of which has trace distance at most
$1-\varepsilon$ from $\left\vert \psi_{n}\right\rangle $, for some polynomial
$p$. \ Because of the Solovay-Kitaev Theorem \cite{kitaev,nc}, $\mathsf{\Psi
P}$\ is invariant under the choice of universal gate set.

\section{Basic Results\label{BASIC}}

Before studying the tree size of specific quantum states, we would like to
know in general how tree size behaves as a complexity measure. \ In this
section we prove three rather nice properties of tree size.

\begin{theorem}
\label{logdepth}For all $\varepsilon>0$, there exists a tree representing
$\left\vert \psi\right\rangle $\ of size $O\left(  \operatorname*{TS}\left(
\left\vert \psi\right\rangle \right)  ^{1+\varepsilon}\right)  $\ and depth
$O\left(  \log\operatorname*{TS}\left(  \left\vert \psi\right\rangle \right)
\right)  $, as well as a manifestly orthogonal tree of size $O\left(
\operatorname*{MOTS}\left(  \left\vert \psi\right\rangle \right)
^{1+\varepsilon}\right)  $\ and depth $O\left(  \log\operatorname*{MOTS}%
\left(  \left\vert \psi\right\rangle \right)  \right)  $.
\end{theorem}

\begin{proof}
A classical theorem of Brent \cite{brent} says that given an arithmetic
formula $\Phi$, there exists an equivalent formula of depth $O\left(
\log\left\vert \Phi\right\vert \right)  $\ and size $O\left(  \left\vert
\Phi\right\vert ^{c}\right)  $, where $c$ is a constant. \ Bshouty, Cleve, and
Eberly \cite{bce}\ (see also Bonet and Buss \cite{bb}) improved Brent's
theorem to show that $c$\ can be taken to be $1+\varepsilon$\ for any
$\varepsilon>0$. \ So it suffices to show that, for `division-free' formulas,
these theorems preserve multilinearity (and in the $\operatorname*{MOTS}%
$\ case, preserve manifest orthogonality).

Brent's theorem is proven by induction on $\left\vert \Phi\right\vert $.
\ Here is a sketch: choose a subformula $I$ of $\Phi$ size between $\left\vert
\Phi\right\vert /3$\ and $2\left\vert \Phi\right\vert /3$\ (which one can show
always exists). \ Then identifying a subformula with the polynomial computed
at its root, $\Phi\left(  x\right)  $\ can be written as $G\left(  x\right)
+H\left(  x\right)  I\left(  x\right)  $\ for some formulas $G$ and $H$.
\ Furthermore, $G$ and $H$\ are both obtainable from $\Phi$ by removing $I$
and then applying further restrictions. \ So $\left\vert G\right\vert $ and
$\left\vert H\right\vert $ are both at most $\left\vert \Phi\right\vert
-\left\vert I\right\vert +O\left(  1\right)  $. \ Let $\widehat{\Phi}$\ be a
formula equivalent to $\Phi$ that evaluates $G$, $H$, and $I$ separately, and
then returns $G\left(  x\right)  +H\left(  x\right)  I\left(  x\right)  $.
\ Then $\left\vert \widehat{\Phi}\right\vert $\ is larger than $\left\vert
\Phi\right\vert $\ by at most a constant factor, while by the induction
hypothesis,\ we can assume the formulas for $G$, $H$, and $I$ have logarithmic
depth. \ Since the number of induction steps is $O\left(  \log\left\vert
\Phi\right\vert \right)  $, the total depth is logarithmic and the total
blowup in formula size is polynomial in $\left\vert \Phi\right\vert $.
\ Bshouty, Cleve, and Eberly's improvement uses a more careful decomposition
of $\Phi$, but the basic idea is the same.

Now, if $\Phi$ is syntactic multilinear, then clearly $G$, $H$, and $I$ are
also syntactic multilinear. \ Furthermore, $H$ cannot share variables with
$I$, since otherwise a subformula of $\Phi$ containing $I$ would have been
multiplied by a subformula containing variables from $I$. \ Thus
multilinearity is preserved. \ To see that manifest orthogonality is
preserved, suppose we are evaluating $G$ and $H$ `bottom up,' and let $G_{v}%
$\ and $H_{v}$\ be the polynomials computed at vertex $v$ of $\Phi$. \ Let
$v_{0}=\operatorname*{root}\left(  I\right)  $, let $v_{1}$\ be the parent of
$v_{0}$, let $v_{2}$\ be the parent of $v_{1}$, and so on until $v_{k}%
=\operatorname*{root}\left(  \Phi\right)  $. \ It is clear that, for every
$x$, either $G_{v_{0}}\left(  x\right)  =0$\ or $H_{v_{0}}\left(  x\right)
=0$. \ Furthermore, suppose that property holds for $G_{v_{i-1}},H_{v_{i-1}}$;
then by induction it holds for $G_{v_{i}},H_{v_{i}}$. \ If $v_{i}$\ is a
$\times$\ gate, then this follows from multilinearity (if $\left\vert
\psi\right\rangle $\ and $\left\vert \varphi\right\rangle $\ are manifestly
orthogonal, then $\left\vert 0\right\rangle \otimes\left\vert \psi
\right\rangle $\ and $\left\vert 0\right\rangle \otimes\left\vert
\varphi\right\rangle $\ are also manifestly orthogonal). \ If $v_{i}$\ is a
$+$ gate, then letting $\operatorname*{supp}\left(  p\right)  $\ be the set of
$x$ such that $p\left(  x\right)  \neq0$, any polynomial $p$ added to
$G_{v_{i-1}}$\ or $H_{v_{i-1}}$\ must have%
\[
\operatorname*{supp}\left(  p\right)  \cap\left(  \operatorname*{supp}\left(
G_{v_{i-1}}\right)  \cup\operatorname*{supp}\left(  H_{v_{i-1}}\right)
\right)  =\emptyset,
\]
and manifest orthogonality follows.
\end{proof}

\begin{theorem}
\label{prepare}Any $\left\vert \psi\right\rangle $\ can be prepared by a
quantum circuit of size polynomial in $\operatorname*{OTS}\left(  \left\vert
\psi\right\rangle \right)  $. \ Thus $\mathsf{OTree}\subseteq\mathsf{\Psi P}$.
\end{theorem}

\begin{proof}
Let $\Gamma\left(  \left\vert \psi\right\rangle \right)  $\ be the minimum
size of a circuit needed to prepare $\left\vert \psi\right\rangle
\in\mathcal{H}_{2}^{\otimes n}$\ starting from $\left\vert 0\right\rangle
^{\otimes n}$.\ \ We prove by induction on $\Gamma\left(  \left\vert
\psi\right\rangle \right)  $ that $\Gamma\left(  \left\vert \psi\right\rangle
\right)  \leq q\left(  \operatorname*{OTS}\left(  \left\vert \psi\right\rangle
\right)  \right)  $ for some polynomial $q$. \ The base case
$\operatorname*{OTS}\left(  \left\vert \psi\right\rangle \right)  =1$\ is
clear. \ Let $T$ be an orthogonal state tree for $\left\vert \psi\right\rangle
$, and assume without loss of generality that every gate has fan-in $2$ (this
increases $\left\vert T\right\vert $ by at most a constant factor). \ Let
$T_{1}$\ and $T_{2}$\ be the subtrees of $\operatorname*{root}\left(
T\right)  $, representing states $\left\vert \psi_{1}\right\rangle $\ and
$\left\vert \psi_{2}\right\rangle $\ respectively; note that $\left\vert
T\right\vert =\left\vert T_{1}\right\vert +\left\vert T_{2}\right\vert $.
\ First suppose $\operatorname*{root}\left(  T\right)  $\ is a $\otimes
$\ gate; then clearly $\Gamma\left(  \left\vert \psi\right\rangle \right)
\leq\Gamma\left(  \left\vert \psi_{1}\right\rangle \right)  +\Gamma\left(
\left\vert \psi_{2}\right\rangle \right)  $.

Second, suppose $\operatorname*{root}\left(  T\right)  $\ is a $+$\ gate, with
$\left\vert \psi\right\rangle =\alpha\left\vert \psi_{1}\right\rangle
+\beta\left\vert \psi_{2}\right\rangle $\ and $\left\langle \psi_{1}|\psi
_{2}\right\rangle =0$. \ Let $U$ be a quantum circuit that prepares
$\left\vert \psi_{1}\right\rangle $, and $V$\ be a circuit\ that prepares
$\left\vert \psi_{2}\right\rangle $.\ \ Then we can prepare $\alpha\left\vert
0\right\rangle \left\vert 0\right\rangle ^{\otimes n}+\beta\left\vert
1\right\rangle U^{-1}V\left\vert 0\right\rangle ^{\otimes n}$. \ Observe that
$U^{-1}V\left\vert 0\right\rangle ^{\otimes n}$\ is orthogonal to $\left\vert
0\right\rangle ^{\otimes n}$, since $\left\vert \psi_{1}\right\rangle
=U\left\vert 0\right\rangle ^{\otimes n}$\ is orthogonal to $\left\vert
\psi_{2}\right\rangle =V\left\vert 0\right\rangle ^{\otimes n}$. \ So applying
a $\operatorname*{NOT}$\ to the first register, conditioned on the
$\operatorname*{OR}$\ of the bits in the second register, yields $\left\vert
0\right\rangle \otimes\left(  \alpha\left\vert 0\right\rangle ^{\otimes
n}+\beta U^{-1}V\left\vert 0\right\rangle ^{\otimes n}\right)  $, from which
we obtain $\alpha\left\vert \psi_{1}\right\rangle +\beta\left\vert \psi
_{2}\right\rangle $\ by applying $U$ to the second register. \ The size of the
circuit used is $O\left(  \left\vert U\right\vert +\left\vert V\right\vert
+n\right)  $, with a possible constant-factor blowup arising from the need to
condition on the first register. \ If we are more careful, however, we can
combine the `conditioning' steps across multiple levels of the recursion,
producing a circuit of size $\left\vert V\right\vert +O\left(  \left\vert
U\right\vert +n\right)  $. \ By symmetry, we can also reverse the roles of $U$
and $V$ to obtain a circuit of size $\left\vert U\right\vert +O\left(
\left\vert V\right\vert +n\right)  $. \ Therefore%
\[
\Gamma\left(  \left\vert \psi\right\rangle \right)  \leq\min\left\{
\Gamma\left(  \left\vert \psi_{1}\right\rangle \right)  +c\Gamma\left(
\left\vert \psi_{2}\right\rangle \right)  +cn,\,\,c\Gamma\left(  \left\vert
\psi_{2}\right\rangle \right)  +\Gamma\left(  \left\vert \psi_{1}\right\rangle
\right)  +cn\right\}
\]
for some constant $c\geq2$. \ Solving this recurrence we find that
$\Gamma\left(  \left\vert \psi\right\rangle \right)  $\ is polynomial in
$\operatorname*{OTS}\left(  \left\vert \psi\right\rangle \right)  $.
\end{proof}

\begin{theorem}
\label{nonapx}If $\left\vert \psi\right\rangle \in\mathcal{H}_{2}^{\otimes n}%
$\ is chosen uniformly at random under the Haar measure, then
$\operatorname*{TS}_{1/16}\left(  \left\vert \psi\right\rangle \right)
=2^{\Omega\left(  n\right)  }$\ with probability $1-o\left(  1\right)  $.
\end{theorem}

\begin{proof}
To generate a uniform random state $\left\vert \psi\right\rangle =\sum
_{x\in\left\{  0,1\right\}  ^{n}}\alpha_{x}\left\vert x\right\rangle $, we can
choose $\widehat{\alpha}_{x},\widehat{\beta}_{x}\in\mathbb{R}$\ for each $x$
independently from a Gaussian distribution with mean $0$ and variance $1$,
then let $\alpha_{x}=\left(  \widehat{\alpha}_{x}+i\widehat{\beta}_{x}\right)
/\sqrt{R}$\ where $R=\sum_{x\in\left\{  0,1\right\}  ^{n}}\left(
\widehat{\alpha}_{x}^{2}+\widehat{\beta}_{x}^{2}\right)  $. \ Let%
\[
\Lambda_{\psi}=\left\{  x:\left(  \operatorname{Re}\alpha_{x}\right)
^{2}<\frac{1}{4\cdot2^{n}}\right\}  ,
\]
and let $\mathcal{G}$\ be the set of $\left\vert \psi\right\rangle $\ for
which $\left\vert \Lambda_{\psi}\right\vert <2^{n}/5$. \ We claim that
$\Pr_{\left\vert \psi\right\rangle }\left[  \left\vert \psi\right\rangle
\in\mathcal{G}\right]  =1-o\left(  1\right)  $. \ First, $\operatorname*{EX}%
\left[  R\right]  =2^{n+1}$, so by a standard Hoeffding-type bound,
$\Pr\left[  R<2^{n}\right]  $\ is doubly-exponentially small in $n$. \ Second,
assuming $R\geq2^{n}$, for each $x$%
\[
\Pr\left[  x\in\Lambda_{\psi}\right]  \leq\Pr\left[  \widehat{\alpha}_{x}%
^{2}<\frac{1}{4}\right]  =\operatorname{erf}\left(  \frac{1}{4\sqrt{2}%
}\right)  <0.198,
\]
and the claim follows by a Chernoff bound.

For $g:\left\{  0,1\right\}  ^{n}\rightarrow\mathbb{R}$, let $A_{g}=\left\{
x:\operatorname*{sgn}\left(  g\left(  x\right)  \right)  \neq
\operatorname*{sgn}\left(  \operatorname{Re}\alpha_{x}\right)  \right\}  $,
where $\operatorname*{sgn}\left(  y\right)  $\ is $1$ if $y\geq0$\ and
$-1$\ otherwise.\ \ Then if $\left\vert \psi\right\rangle \in\mathcal{G}$,
clearly%
\[
\sum_{x\in\left\{  0,1\right\}  ^{n}}\left\vert g\left(  x\right)  -f_{\psi
}\left(  x\right)  \right\vert ^{2}\geq\frac{\left\vert A_{g}\right\vert
-\left\vert \Lambda_{\psi}\right\vert }{4\cdot2^{n}}%
\]
where $f_{\psi}\left(  x\right)  =\operatorname{Re}\alpha_{x}$, and thus%
\[
\left\vert A_{g}\right\vert \leq\left(  4\left\Vert g-f_{\psi}\right\Vert
_{2}^{2}+\frac{1}{5}\right)  2^{n}.
\]
Therefore to show that $\operatorname*{MFS}_{1/15}\left(  f_{\psi}\right)
=2^{\Omega\left(  n\right)  }$\ with probability $1-o\left(  1\right)  $, we
need only show that for almost all Boolean functions $f:\left\{  0,1\right\}
^{n}\rightarrow\left\{  -1,1\right\}  $, there is no arithmetic formula $\Phi$
of size $2^{o\left(  n\right)  }$\ such that
\[
\left\vert \left\{  x:\operatorname*{sgn}\left(  \Phi\left(  x\right)
\right)  \neq f\left(  x\right)  \right\}  \right\vert \leq0.49\cdot2^{n}.
\]
Here an arithmetic formula is real-valued, and can include addition,
subtraction, and multiplication gates of fan-in $2$ as well as constants. \ We
do not need to assume multilinearity, and it is easy to see that the
assumption of bounded fan-in is without loss of generality. \ Let $W$ be the
set of Boolean functions \textit{sign-represented} by an arithmetic formula
$\Phi$ of size $2^{o\left(  n\right)  }$, in the sense that
$\operatorname*{sgn}\left(  \Phi\left(  x\right)  \right)  =f\left(  x\right)
$ for all $x$. \ Then it suffices to show that $\left\vert W\right\vert
=2^{2^{o\left(  n\right)  }}$, since the number of functions sign-represented
on an $0.51$\ fraction of inputs is at most $\left\vert W\right\vert
\cdot2^{2^{n}H\left(  0.51\right)  }$.\ \ (Here $H$ denotes the binary entropy function.)

Let $\Phi$ be an arithmetic formula that takes as input the binary string
$x=\left(  x_{1},\ldots,x_{n}\right)  $\ as well as constants $c_{1}%
,c_{2},\ldots$. \ Let $\Phi_{c}$\ denote $\Phi$ under a particular assignment
$c$ to $c_{1},c_{2},\ldots$.\ \ Then a result of Gashkov \cite{gashkov} (see
also Tur\'{a}n and Vatan \cite{tv}), which follows from Warren's Theorem
\cite{warren}\ in real algebraic geometry, shows that as we range over all
$c$, $\Phi_{c}$ sign-represents at most $\left(  2^{n+4}\left\vert
\Phi\right\vert \right)  ^{\left\vert \Phi\right\vert }$\ distinct Boolean
functions, where $\left\vert \Phi\right\vert $\ is the size of $\Phi$.
\ Furthermore, excluding constants, the number of distinct arithmetic formulas
of size $\left\vert \Phi\right\vert $\ is at most $\left(  3\left\vert
\Phi\right\vert ^{2}\right)  ^{\left\vert \Phi\right\vert }$. \ When
$\left\vert \Phi\right\vert =2^{o\left(  n\right)  }$, this gives $\left(
3\left\vert \Phi\right\vert ^{2}\right)  ^{\left\vert \Phi\right\vert }%
\cdot\left(  2^{n+4}\left\vert \Phi\right\vert \right)  ^{\left\vert
\Phi\right\vert }=2^{2^{o\left(  n\right)  }}$. \ We have shown that
$\operatorname*{MFS}_{1/15}\left(  f_{\psi}\right)  =2^{\Omega\left(
n\right)  }$; by Theorem \ref{iff}, part (iii), this implies that
$\operatorname*{TS}_{1/16}\left(  \left\vert \psi\right\rangle \right)
=2^{\Omega\left(  n\right)  }$.
\end{proof}

A corollary of Theorem \ref{nonapx} is the following `nonamplification'
property: there exist states that can be approximated to within, say,
$1\%$\ by trees of polynomial size, but that require exponentially large trees
to approximate to within a smaller margin (say $0.01\%$).

\begin{corollary}
\label{nonamp}For all $\delta\in\left(  0,1\right]  $, there exists a state
$\left\vert \psi\right\rangle $\ such that $\operatorname*{TS}_{\delta}\left(
\left\vert \psi\right\rangle \right)  =n$\ but $\operatorname*{TS}%
_{\varepsilon}\left(  \left\vert \psi\right\rangle \right)  =2^{\Omega\left(
n\right)  }$ where $\varepsilon=\delta/32-\delta^{2}/4096$.
\end{corollary}

\begin{proof}
It is clear from Theorem \ref{nonapx}\ that there exists a state $\left\vert
\varphi\right\rangle =\sum_{x\in\left\{  0,1\right\}  ^{n}}\alpha
_{x}\left\vert x\right\rangle $\ such that $\operatorname*{TS}_{1/16}\left(
\left\vert \varphi\right\rangle \right)  =2^{\Omega\left(  n\right)  }$\ and
$\alpha_{0^{n}}=0$. \ Take $\left\vert \psi\right\rangle =\sqrt{1-\delta
}\left\vert 0\right\rangle ^{\otimes n}+\sqrt{\delta}\left\vert \varphi
\right\rangle $. \ Since $\left\vert \left\langle \psi|0\right\rangle
^{\otimes n}\right\vert ^{2}=1-\delta$, we have $\operatorname*{MOTS}_{\delta
}\left(  \left\vert \psi\right\rangle \right)  =n$. \ On the other hand,
suppose some $\left\vert \phi\right\rangle =\sum_{x\in\left\{  0,1\right\}
^{n}}\beta_{x}\left\vert x\right\rangle $\ with $\operatorname*{TS}\left(
\left\vert \phi\right\rangle \right)  =2^{o\left(  n\right)  }$ satisfies
$\left\vert \left\langle \phi|\psi\right\rangle \right\vert ^{2}%
\geq1-\varepsilon$. \ Then%
\[
\sum_{x\neq0^{n}}\left(  \sqrt{\delta}\alpha_{x}-\beta_{x}\right)  ^{2}%
\leq2-2\sqrt{1-\varepsilon}.
\]
Thus, letting $f_{\varphi}\left(  x\right)  =\alpha_{x}$, we have
$\operatorname*{MFS}\nolimits_{c}\left(  f_{\varphi}\right)  =O\left(
\operatorname*{TS}\left(  \left\vert \phi\right\rangle \right)  \right)
$\ where $c=\left(  2-2\sqrt{1-\varepsilon}\right)  /\delta$. \ By Theorem
\ref{iff}, part (iv), this implies that $\operatorname*{TS}_{2c}\left(
\left\vert \varphi\right\rangle \right)  =O\left(  \operatorname*{TS}\left(
\left\vert \phi\right\rangle \right)  \right)  $. \ But $2c=1/16$\ when
$\varepsilon=\delta/32-\delta^{2}/4096$, contradiction.
\end{proof}

\section{Lower Bounds\label{LOWER}}

We want to show that certain quantum states of interest to us are not
represented by trees of polynomial size. \ At first this seems like a hopeless
task. \ Proving superpolynomial formula-size lower bounds for `explicit'
functions is a notoriously hard open problem, as it would imply complexity
class separations such as $\mathsf{NC}^{1}\neq\mathsf{P}$.

Here, though, we are only concerned with \textit{multilinear} formulas.
\ Could this make it easier to prove a lower bound? \ The answer is not
obvious, but very recently, for reasons unrelated to quantum computing, Raz
\cite{raz,raz2}\ showed the first superpolynomial lower bounds on multilinear
formula size. \ In particular, he showed that multilinear formulas computing
the permanent or determinant of an $n\times n$\ matrix over any field have
size $n^{\Omega\left(  \log n\right)  }$.

Raz's technique is a beautiful combination of the Furst-Saxe-Sipser method of
random restrictions \cite{fss}, with matrix rank arguments as used in
communication complexity. \ We now outline the method. \ Given a function
$f:\left\{  0,1\right\}  ^{n}\rightarrow\mathbb{C}$, let $P$ be a partition of
the input variables $x_{1},\ldots,x_{n}$\ into two collections $y=\left(
y_{1},\ldots,y_{n/2}\right)  $\ and $z=\left(  z_{1},\ldots,z_{n/2}\right)  $.
\ This yields a function $f_{P}\left(  y,z\right)  :\left\{  0,1\right\}
^{n/2}\times\left\{  0,1\right\}  ^{n/2}\rightarrow\mathbb{C}$. \ Then let
$M_{f|P}$\ be a $2^{n/2}\times2^{n/2}$\ matrix whose rows are labeled by
assignments $y\in\left\{  0,1\right\}  ^{n/2}$, and whose columns are labeled
by assignments $z\in\left\{  0,1\right\}  ^{n/2}$. \ The $\left(  y,z\right)
$\ entry of $M_{f|P}$\ is $f_{P}\left(  y,z\right)  $. \ Let
$\operatorname*{rank}\left(  M_{f|P}\right)  $\ be the rank of $M_{f|P}$ over
the complex numbers. \ Finally, let $\mathcal{P}$\ be the uniform distribution
over all partitions $P$.

The following, Corollary 3.6 in \cite{raz2}, is one statement of Raz's main
theorem; recall that $\operatorname*{MFS}\left(  f\right)  $\ is the minimum
size of a multilinear formula for $f$.

\begin{theorem}
[\cite{raz2}]\label{razthm}Suppose that%
\[
\Pr_{P\in\mathcal{P}}\left[  \operatorname*{rank}\left(  M_{f|P}\right)
\geq2^{n/2-\left(  n/2\right)  ^{1/8}/2}\right]  =n^{-o\left(  \log n\right)
}.
\]
Then $\operatorname*{MFS}\left(  f\right)  =n^{\Omega\left(  \log n\right)  }$.
\end{theorem}

An immediate corollary yields lower bounds on \textit{approximate} multilinear
formula size. \ Given an $N\times N$ matrix $M=\left(  m_{ij}\right)  $, let
$\operatorname*{rank}\nolimits_{\varepsilon}\left(  M\right)  =\min
_{L~:~\left\Vert L-M\right\Vert _{2}^{2}\leq\varepsilon}\operatorname*{rank}%
\left(  L\right)  $\ where $\left\Vert L-M\right\Vert _{2}^{2}=\sum
_{i,j=1}^{N}\left\vert l_{ij}-m_{ij}\right\vert ^{2}$.

\begin{corollary}
\label{razcor}Suppose that%
\[
\Pr_{P\in\mathcal{P}}\left[  \operatorname*{rank}\nolimits_{\varepsilon
}\left(  M_{f|P}\right)  \geq2^{n/2-\left(  n/2\right)  ^{1/8}/2}\right]
=n^{-o\left(  \log n\right)  }.
\]
Then $\operatorname*{MFS}_{\varepsilon}\left(  f\right)  =n^{\Omega\left(
\log n\right)  }$.
\end{corollary}

\begin{proof}
Suppose $\operatorname*{MFS}_{\varepsilon}\left(  f\right)  =n^{o\left(  \log
n\right)  }$. \ Then for all $g$ such that $\left\Vert f-g\right\Vert _{2}%
^{2}\leq\varepsilon$, we would have $\operatorname*{MFS}\left(  g\right)
=n^{o\left(  \log n\right)  }$, and therefore%
\[
\Pr_{P\in\mathcal{P}}\left[  \operatorname*{rank}\left(  M_{g|P}\right)
\geq2^{n/2-\left(  n/2\right)  ^{1/8}/2}\right]  =n^{-\Omega\left(  \log
n\right)  }.
\]
by Theorem \ref{razthm}. \ But $\operatorname*{rank}\nolimits_{\varepsilon
}\left(  M_{f|P}\right)  \leq\operatorname*{rank}\left(  M_{g|P}\right)  $,
and hence%
\[
\Pr_{P\in\mathcal{P}}\left[  \operatorname*{rank}\nolimits_{\varepsilon
}\left(  M_{f|P}\right)  \geq2^{n/2-\left(  n/2\right)  ^{1/8}/2}\right]
=n^{-\Omega\left(  \log n\right)  },
\]
contradiction.
\end{proof}

Another simple corollary gives lower bounds in terms of \textit{restrictions}
of $f$. \ Let $\mathcal{R}_{l}$ be the following distribution over
restrictions $R$: choose $2l$\ variables of $f$ uniformly at random, and
rename them $y=\left(  y_{1},\ldots,y_{l}\right)  $\ and $z=\left(
z_{1},\ldots,z_{l}\right)  $. \ Set each of the remaining $n-2l$\ variables to
$0$\ or $1$ uniformly and independently at random. \ This yields a restricted
function $f_{R}\left(  y,z\right)  $. \ Let $M_{f|R}$\ be a $2^{l}\times2^{l}%
$\ matrix whose $\left(  y,z\right)  $\ entry is $f_{R}\left(  y,z\right)  $.

\begin{corollary}
\label{razcor2}Suppose that%
\[
\Pr_{R\in\mathcal{R}_{l}}\left[  \operatorname*{rank}\left(  M_{f|R}\right)
\geq2^{l-l^{1/8}/2}\right]  =n^{-o\left(  \log n\right)  }%
\]
where $l=n^{\delta}$\ for some constant $\delta\in\left(  0,1\right]  $.
\ Then $\operatorname*{MFS}\left(  f\right)  =n^{\Omega\left(  \log n\right)
}$.
\end{corollary}

\begin{proof}
Under the hypothesis, clearly there exists a \textit{fixed} restriction
$g:\left\{  0,1\right\}  ^{2l}\rightarrow\mathbb{C}$\ of $f$, which leaves
$2l$\ variables unrestricted, such that%
\[
\Pr_{P\in\mathcal{P}}\left[  \operatorname*{rank}\left(  M_{g|P}\right)
\geq2^{l-l^{1/8}/2}\right]  =n^{-o\left(  \log n\right)  }=l^{-o\left(  \log
l\right)  }.
\]
Then by Theorem \ref{razthm},%
\[
\operatorname*{MFS}\left(  f\right)  \geq\operatorname*{MFS}\left(  g\right)
=l^{\Omega\left(  \log l\right)  }=n^{\Omega\left(  \log n\right)  }.
\]

\end{proof}

We will apply Raz's theorem to obtain $n^{\Omega\left(  \log n\right)  }%
$\ tree size\ lower bounds\ for two classes of quantum states: states arising
in quantum error-correction in Section \ref{ECC}, and (assuming a
number-theoretic conjecture) states arising in Shor's factoring algorithm in
Section \ref{DIVIS}.

\subsection{Subgroup States\label{ECC}}

Let the elements of $\mathbb{Z}_{2}^{n}$\ be labeled by $n$-bit strings.
\ Given a subgroup $S\leq\mathbb{Z}_{2}^{n}$, we define the \textit{subgroup
state} $\left\vert S\right\rangle $\ as follows:%
\[
\left\vert S\right\rangle =\frac{1}{\sqrt{\left\vert S\right\vert }}\sum_{x\in
S}\left\vert x\right\rangle .
\]
Coset states arise as codewords in the class of quantum error-correcting codes
known as stabilizer codes \cite{cs,gottesman,steane}. \ Our interest in these
states, however, arises from their large tree size rather than their
error-correcting properties.

Let $\mathcal{E}$\ be the following distribution over subgroups $S$. \ Choose
an $n/2\times n$\ matrix $A$ by setting each entry to $0$ or $1$ uniformly and
independently. \ Then let $S=\left\{  x~|~Ax\equiv0\left(  \operatorname{mod}%
2\right)  \right\}  $. \ By Theorem \ref{iff}, part (i), it suffices to
lower-bound the multilinear formula size of the function $f_{S}\left(
x\right)  $, which is $1$ if $x\in S$\ and $0$\ otherwise.

\begin{theorem}
\label{ecclb}If $S$ is drawn from $\mathcal{E}$, then $\operatorname*{MFS}%
\left(  f_{S}\right)  =n^{\Omega\left(  \log n\right)  }$ (and hence
$\operatorname*{TS}\left(  \left\vert S\right\rangle \right)  =n^{\Omega
\left(  \log n\right)  }$), with probability $\Omega\left(  1\right)  $\ over
$S$.
\end{theorem}

\begin{proof}
Let $P$ be a uniform random partition of the inputs $x_{1},\ldots,x_{n}$\ of
$f_{S}$\ into two sets $y=\left(  y_{1},\ldots,y_{n/2}\right)  $\ and
$z=\left(  z_{1},\ldots,z_{n/2}\right)  $. \ Let $M_{S|P}$\ be the
$2^{n/2}\times2^{n/2}$\ matrix whose $\left(  y,z\right)  $\ entry is
$f_{S|P}\left(  y,z\right)  $; then we need to show that $\operatorname*{rank}%
\left(  M_{S|P}\right)  $\ is large with high probability. \ Let $A_{y}$\ be
the $n/2\times n/2$\ submatrix of the $n/2\times n$\ matrix $A$ consisting of
all rows that correspond to $y_{i}$\ for some $i\in\left\{  1,\ldots
,n/2\right\}  $,\ and similarly let $A_{z}$\ be the $n/2\times n/2$%
\ submatrix\ corresponding to $z$. \ Then it is easy to see that, so long as
$A_{y}$\ and $A_{z}$\ are both invertible, for all $2^{n/2}$\ settings of $y$
there exists a \textit{unique} setting of $z$ for which $f_{S|P}\left(
y,z\right)  =1$. \ This then implies that $M_{S|P}$\ is a permutation of the
identity matrix, and hence that $\operatorname*{rank}\left(  M_{S|P}\right)
=2^{n/2}$. \ Now, the probability that a random $n/2\times n/2$\ matrix over
$\mathbb{Z}_{2}$\ is invertible is%
\[
\frac{1}{2}\cdot\frac{3}{4}\cdot\cdots\cdot\frac{2^{n/2}-1}{2^{n/2}}>0.288.
\]
So the probability that $A_{y}$ and $A_{z}$ are both invertible is at least
$0.288^{2}$. \ By Markov's inequality, it follows that for at least an $0.04$
fraction of $S$'s, $\operatorname*{rank}\left(  M_{S|P}\right)  =2^{n/2}$\ for
at least an $0.04$\ fraction of $P$'s. \ Theorem \ref{razthm}\ then yields the
desired result.
\end{proof}

Aaronson and Gottesman \cite{ag}\ show how to prepare any $n$-qubit subgroup
state using a quantum circuit of size $O\left(  n^{2}/\log n\right)  $. \ So a
corollary of Theorem \ref{ecclb} is that $\mathsf{\Psi P}\not \subset
\mathsf{Tree}$. \ Since $f_{S}$ clearly has a (non-multilinear) arithmetic
formula of size $O\left(  nk\right)  $, a second corollary is the following.

\begin{corollary}
\label{mlinsep}There exists a family of functions $f_{n}:\left\{  0,1\right\}
^{n}\rightarrow\mathbb{R}$\ that has polynomial-size arithmetic formulas, but
no polynomial-size multilinear formulas.
\end{corollary}

The reason Corollary \ref{mlinsep}\ does not follow from Raz's results is that
polynomial-size formulas for the permanent and determinant are not known; the
smallest known formulas for the determinant have size $n^{O\left(  \log
n\right)  }$\ (see \cite{bcs}).

We have shown that not all subgroup states are tree states, but it is still
conceivable that all subgroup states are extremely well \textit{approximated}
by tree states. \ Let us now rule out the latter possibility. \ We first need
a lemma about matrix rank, which follows from the Hoffman-Wielandt inequality.

\begin{lemma}
\label{hw}Let $M$ be an $N\times N$\ complex matrix, and let $I_{N}$\ be the
$N\times N$\ identity matrix. \ Then $\left\Vert M-I_{N}\right\Vert _{2}%
^{2}\geq N-\operatorname*{rank}\left(  M\right)  $.
\end{lemma}

\begin{proof}
The Hoffman-Wielandt inequality \cite{hw}\ (see also \cite{astvw}) states that
for any two $N\times N$\ matrices $M,P$,%
\[
\sum_{i=1}^{N}\left(  \sigma_{i}\left(  M\right)  -\sigma_{i}\left(  P\right)
\right)  ^{2}\leq\left\Vert M-P\right\Vert _{2}^{2},
\]
where $\sigma_{i}\left(  M\right)  $\ is the $i^{th}$\ singular value of
$M$\ (that is, $\sigma_{i}\left(  M\right)  =\sqrt{\lambda_{i}\left(
M\right)  }$, where $\lambda_{1}\left(  M\right)  \geq\cdots\geq\lambda
_{N}\left(  M\right)  \geq0$\ are the eigenvalues of $MM^{\ast}$, and
$M^{\ast}$\ is the conjugate transpose of $M$). \ Clearly $\sigma_{i}\left(
I_{N}\right)  =1$\ for all $i$. \ On the other hand, $M$ has only
$\operatorname*{rank}\left(  M\right)  $\ nonzero singular values, so%
\[
\sum_{i=1}^{N}\left(  \sigma_{i}\left(  M\right)  -\sigma_{i}\left(
I_{N}\right)  \right)  ^{2}\geq N-\operatorname*{rank}\left(  M\right)  .
\]

\end{proof}

Let $\widehat{f}_{S}\left(  x\right)  =f_{S}\left(  x\right)  /\sqrt
{\left\vert S\right\vert }$ be $f_{S}\left(  x\right)  $ normalized to have
$\left\Vert \widehat{f}_{S}\right\Vert _{2}^{2}=1$.

\begin{theorem}
\label{ecclbapprox}For all constants $\varepsilon\in\left[  0,1\right)  $, if
$S$ is drawn from $\mathcal{E}$, then $\operatorname*{MFS}_{\varepsilon
}\left(  \widehat{f}_{S}\right)  =n^{\Omega\left(  \log n\right)  }$ with
probability $\Omega\left(  1\right)  $\ over $S$.
\end{theorem}

\begin{proof}
As in Theorem \ref{ecclb}, we look at the matrix $M_{S|P}$\ induced by a
random partition $P=\left(  y,z\right)  $. \ We already know that for at least
an $0.04$\ fraction of $S$'s, the $y$ and $z$ variables are in one-to-one
correspondence for at least an $0.04$\ fraction of $P$'s. \ In that case
$\left\vert S\right\vert =2^{n/2}$, and therefore $M_{S|P}$\ is a permutation
of $I/\sqrt{\left\vert S\right\vert }=I/2^{n/4}$\ where $I$ is the identity.
\ It follows from Lemma \ref{hw}\ that for all matrices $M$ such that
$\left\Vert M-M_{S|P}\right\Vert _{2}^{2}\leq\varepsilon$,%
\[
\operatorname*{rank}\left(  M\right)  \geq2^{n/2}-\left\Vert \sqrt{\left\vert
S\right\vert }\left(  M-M_{S|P}\right)  \right\Vert _{2}^{2}\geq\left(
1-\varepsilon\right)  2^{n/2}%
\]
and therefore $\operatorname*{rank}\nolimits_{\varepsilon}\left(
M_{S|P}\right)  \geq\left(  1-\varepsilon\right)  2^{n/2}$. \ Hence%
\[
\Pr_{P\in\mathcal{P}}\left[  \operatorname*{rank}\nolimits_{\varepsilon
}\left(  M_{f|P}\right)  \geq2^{n/2-\left(  n/2\right)  ^{1/8}/2}\right]
\geq0.04,
\]
and the result follows from Corollary \ref{razcor}.
\end{proof}

A corollary of Theorem \ref{ecclbapprox}\ and of Theorem \ref{iff}, part
(iii), is that $\operatorname*{TS}_{\varepsilon}\left(  \left\vert
S\right\rangle \right)  =n^{\Omega\left(  \log n\right)  }$ with probability
$\Omega\left(  1\right)  $\ over $S$, for all $\varepsilon<1$.

Finally, let us show how to derandomize the lower bound for subgroup states,
using ideas pointed out to us by Andrej Bogdanov. \ In the proof of Theorem
\ref{ecclb}, all we used about the matrix $A$ was that a random $k\times
k$\ submatrix has full rank with $\Omega\left(  1\right)  $ probability, where
$k=n/2$. \ If we switch from the field $\mathbb{F}_{2}$\ to $\mathbb{F}%
_{2^{d}}$ for some $d\geq\log_{2}n$, then it is easy to construct explicit
$k\times n$\ matrices with this same property. \ For example, let%
\[
V=\left(
\begin{array}
[c]{cccc}%
1^{0} & 1^{1} & \cdots & 1^{k-1}\\
2^{0} & 2^{1} & \cdots & 2^{k-1}\\
\vdots & \vdots &  & \vdots\\
n^{0} & n^{1} & \cdots & n^{k-1}%
\end{array}
\right)
\]
be the $n\times k$ Vandermonde matrix, where $1,\ldots,n$\ are labels of
elements in $\mathbb{F}_{2^{d}}$. \ \textit{Any} $k\times k$\ submatrix of $V$
has full rank, because the Reed-Solomon (RS) code that $V$ represents is a
perfect erasure code.\footnote{In other words, because a degree-$\left(
k-1\right)  $ polynomial is determined by its values at any $k$ points.}
\ Hence, there exists an explicit state of $n$ \textquotedblleft
qupits\textquotedblright\ with $p=2^{d}$\ that has tree size $n^{\Omega\left(
\log n\right)  }$---namely the uniform superposition over all elements of the
set $\left\{  x~|~V^{T}x=0\right\}  $, where $V^{T}$\ is the transpose of $V$.

To replace qupits by qubits, we concatenate the RS and Hadamard codes to
obtain a \textit{binary} linear erasure code with parameters almost as good as
those of the original RS code. \ More explicitly, interpret $\mathbb{F}%
_{2^{d}}$\ as the field of polynomials over $\mathbb{F}_{2}$, modulo some
irreducible of degree $d$. \ Then let $m\left(  a\right)  $\ be the $d\times
d$\ Boolean matrix that maps $q\in\mathbb{F}_{2^{d}}$\ to $aq\in
\mathbb{F}_{2^{d}}$, where $q$\ and $aq$\ are encoded by their $d\times
1$\ vectors of coefficients. \ Let $H$ map a length-$d$ vector to its
length-$2^{d}$ Hadamard encoding. \ Then $Hm\left(  a\right)  $\ is a
$2^{d}\times d$\ Boolean matrix that maps $q\in\mathbb{F}_{2^{d}}$\ to the
Hadamard encoding of $aq$. \ We can now define an $n2^{d}\times kd$%
\ \textquotedblleft binary Vandermonde matrix\textquotedblright\ as follows:%
\[
V_{\operatorname*{bin}}=\left(
\begin{array}
[c]{cccc}%
Hm\left(  1^{0}\right)  & Hm\left(  1^{1}\right)  & \cdots & Hm\left(
1^{k-1}\right) \\
Hm\left(  2^{0}\right)  & Hm\left(  2^{1}\right)  & \cdots & Hm\left(
2^{k-1}\right) \\
\vdots & \vdots &  & \vdots\\
Hm\left(  n^{0}\right)  & Hm\left(  n^{1}\right)  & \cdots & Hm\left(
n^{k-1}\right)
\end{array}
\right)  .
\]
For the remainder of the section, fix $k=n^{\delta}$\ for some $\delta
<1/2$\ and $d=O\left(  \log n\right)  $.

\begin{lemma}
\label{kd}A $\left(  kd+c\right)  \times kd$\ submatrix of
$V_{\operatorname*{bin}}$\ chosen uniformly at random has rank $kd$\ (that is,
full rank) with probability at least $2/3$, for $c$ a sufficiently large constant.
\end{lemma}

\begin{proof}
We claim that $\left\vert V_{\operatorname*{bin}}u\right\vert \geq\left(
n-k\right)  2^{d-1}$\ for all nonzero vectors $u\in\mathbb{F}_{2}^{kd}$, where
$\left\vert ~~\right\vert $\ represents the number of `$1$'\ bits. \ To see
this, observe that for all nonzero $u$, the \textquotedblleft codeword
vector\textquotedblright\ $Vu\in\mathbb{F}_{2^{d}}^{n}$\ must have at least
$n-k$ nonzero entries by the Fundamental Theorem of Algebra, where here $u$ is
interpreted as an element of $\mathbb{F}_{2^{d}}^{k}$. \ Furthermore, the
Hadamard code maps any nonzero entry in $Vu$\ to $2^{d-1}$\ nonzero bits in
$V_{\operatorname*{bin}}u\in\mathbb{F}_{2}^{n2^{d}}$.

Now let $W$\ be a uniformly random $\left(  kd+c\right)  \times kd$\ submatrix
of $V_{\operatorname*{bin}}$. \ By the above claim, for any fixed nonzero
vector $u\in\mathbb{F}_{2}^{kd}$,%
\[
\Pr_{W}\left[  Wu=0\right]  \leq\left(  1-\frac{\left(  n-k\right)  2^{d-1}%
}{n2^{d}}\right)  ^{kd+c}=\left(  \frac{1}{2}+\frac{k}{2n}\right)  ^{kd+c}.
\]
So by the union bound, $Wu$\ is nonzero for all nonzero $u$ (and hence $W$ is
full rank) with probability at least%
\[
1-2^{kd}\left(  \frac{1}{2}+\frac{k}{2n}\right)  ^{kd+c}=1-\left(  1+\frac
{k}{n}\right)  ^{kd}\left(  \frac{1}{2}+\frac{k}{2n}\right)  ^{c}.
\]
Since $k=n^{1/2-\Omega\left(  1\right)  }$\ and $d=O\left(  \log n\right)  $,
the above quantity is at least $2/3$\ for sufficiently large $c$.
\end{proof}

Given an $n2^{d}\times1$\ Boolean vector $x$, let $f\left(  x\right)  =1$\ if
$V_{\operatorname*{bin}}^{T}x=0$\ and $f\left(  x\right)  =0$\ otherwise. \ Then:

\begin{theorem}
\label{explicit}$\operatorname*{MFS}\left(  f\right)  =n^{\Omega\left(  \log
n\right)  }$.
\end{theorem}

\begin{proof}
Let $V_{y}$\ and $V_{z}$ be two disjoint $kd\times\left(  kd+c\right)
$\ submatrices of $V_{\operatorname*{bin}}^{T}$ chosen uniformly at random.
\ Then by Lemma \ref{kd}\ together with the union bound, $V_{y}$\ and $V_{z}%
$\ both have full rank with probability at least $1/3$. \ Letting $l=kd+c$, it
follows that%
\[
\Pr_{R\in\mathcal{R}_{l}}\left[  \operatorname*{rank}\left(  M_{f|R}\right)
\geq2^{l-c}\right]  \geq\frac{1}{3}=n^{-o\left(  \log n\right)  }%
\]
by the same reasoning as in Theorem \ref{ecclb}. \ Therefore
$\operatorname*{MFS}\left(  f\right)  =n^{\Omega\left(  \log n\right)  }$\ by
Corollary \ref{razcor2}.
\end{proof}

Let $\left\vert S\right\rangle $\ be a uniform superposition over all $x$ such
that $f\left(  x\right)  =1$; then a corollary of Theorem \ref{explicit}\ is
that $\operatorname*{TS}\left(  \left\vert S\right\rangle \right)
=n^{\Omega\left(  \log n\right)  }$. \ Naturally, using the ideas of
Theorem\ \ref{ecclbapprox} one can also show that $\operatorname*{TS}%
_{\varepsilon}\left(  \left\vert S\right\rangle \right)  =n^{\Omega\left(
\log n\right)  }$\ for all $\varepsilon<1$.

\subsection{Shor States\label{DIVIS}}

Since the motivation for our theory was to study possible Sure/Shor
separators, an obvious question is, \textit{do states arising in Shor's
algorithm have superpolynomial tree size?} \ Unfortunately, we are only able
to answer this question assuming a number-theoretic conjecture. \ To formalize
the question,\ let%
\[
\frac{1}{2^{n/2}}\sum_{r=0}^{2^{n}-1}\left\vert r\right\rangle \left\vert
x^{r}\operatorname{mod}N\right\rangle
\]
be a Shor state. \ It will be convenient for us to measure the second
register, so that the state of the first register has the form%
\[
\left\vert a+p\mathbb{Z}\right\rangle =\frac{1}{\sqrt{I}}\sum_{i=0}%
^{I}\left\vert a+pi\right\rangle
\]
for some integers $a<p$\ and $I=\left\lfloor \left(  2^{n}-a-1\right)
/p\right\rfloor $. \ Here $a+pi$\ is written out in binary using $n$ bits.
\ Clearly a lower bound on $\operatorname*{TS}\left(  \left\vert
a+p\mathbb{Z}\right\rangle \right)  $\ would imply an equivalent lower bound
for the joint state of the two registers. \ Also, to avoid some technicalities
we assume $p$ is prime. \ Since our goal is to prove a \textit{lower} bound,
this assumption is without loss of generality.

Given an $n$-bit string $x=x_{n-1}\ldots x_{0}$, let $f_{n,p,a}\left(
x\right)  =1$\ if $x\equiv a\left(  \operatorname{mod}p\right)  $\ and
$f_{n,p,a}\left(  x\right)  =0$\ otherwise. \ Then $\operatorname*{TS}\left(
\left\vert a+p\mathbb{Z}\right\rangle \right)  =\Theta\left(
\operatorname*{MFS}\left(  f_{n,p,a}\right)  \right)  $\ by Theorem \ref{iff},
so from now on we will focus attention on $f_{n,p,a}$.

\begin{proposition}
\label{trivshor}\quad

\begin{enumerate}
\item[(i)] Let $f_{n,p}=f_{n,p,0}$. \ Then$\ \operatorname*{MFS}\left(
f_{n,p,a}\right)  \leq\operatorname*{MFS}\left(  f_{n+\log p,p}\right)  $,
meaning that we can set $a=0$ without loss of generality.

\item[(ii)] $\operatorname*{MFS}\left(  f_{n,p}\right)  =O\left(
\operatorname*{min}\left\{  n2^{n}/p,np\right\}  \right)  $.
\end{enumerate}
\end{proposition}

\begin{proof}
\quad

\begin{enumerate}
\item[(i)] Take the formula for $f_{n+\log p,p}$, and restrict the most
significant $\log p$\ bits to sum to a number congruent to
$-a\operatorname{mod}p$ (this is always possible since $x\rightarrow2^{n}%
x$\ is an isomorphism of $\mathbb{Z}_{p}$).

\item[(ii)] For $\operatorname*{MFS}\left(  f_{n,p}\right)  =O\left(
n2^{n}/p\right)  $, write out the $x$'s for which $f_{n,p}\left(  x\right)
=1$\ explicitly. \ For $\operatorname*{MFS}\left(  f_{n,p}\right)  =O\left(
np\right)  $, use the Fourier transform, similarly to Theorem \ref{trivrelate}%
, part (v):%
\[
f_{n,p}\left(  x\right)  =\frac{1}{p}\sum_{h=0}^{p-1}%
{\displaystyle\prod\limits_{j=0}^{n-1}}
\exp\left(  \frac{2\pi ih}{p}\cdot2^{j}x_{j}\right)  .
\]
This immediately yields a sum-of-products formula of size $O\left(  np\right)
$.
\end{enumerate}
\end{proof}

We now state our number-theoretic conjecture.

\begin{conjecture}
\label{primes}There exist constants $\gamma,\delta\in\left(  0,1\right)
$\ and a prime $p=\Omega\left(  2^{n^{\delta}}\right)  $ for which the
following holds. \ Let the set $A$ consist of $n^{\delta}$\ elements of
$\left\{  2^{0},\ldots,2^{n-1}\right\}  $ chosen uniformly at random. \ Let
$S$ consist of all $2^{n^{\delta}}$ sums of subsets of $A$,\ and let
$S\operatorname{mod}p=\left\{  x\operatorname{mod}p:x\in S\right\}  $. \ Then%
\[
\Pr_{A}\left[  \left\vert S\operatorname{mod}p\right\vert \geq\left(
1+\gamma\right)  \frac{p}{2}\right]  =n^{-o\left(  \log n\right)  }.
\]

\end{conjecture}

\begin{theorem}
\label{conjimp}Conjecture \ref{primes}\ implies that\ $\operatorname*{MFS}%
\left(  f_{n,p}\right)  =n^{\Omega\left(  \log n\right)  }$\ and hence
$\operatorname*{TS}\left(  \left\vert p\mathbb{Z}\right\rangle \right)
=n^{\Omega\left(  \log n\right)  }$.
\end{theorem}

\begin{proof}
Let $f=f_{n,p}$ and $l=n^{\delta}$. \ Let $R$\ be a restriction of $f$ that
renames $2l$\ variables $y_{1},\ldots,y_{l},z_{1},\ldots,z_{l}$, and sets each
of the remaining $n-2l$\ variables to $0$\ or $1$. \ This leads to a new
function, $f_{R}\left(  y,z\right)  $, which is $1$ if $y+z+c\equiv0\left(
\operatorname{mod}p\right)  $\ and $0$ otherwise for some constant $c$. \ Here
we are defining $y=2^{a_{1}}y_{1}+\cdots+2^{a_{l}}y_{l}$\ and $z=2^{b_{1}%
}z_{1}+\cdots+2^{b_{l}}z_{l}$\ where $a_{1},\ldots,a_{l},b_{1},\ldots,b_{l}%
$\ are the appropriate place values. \ Now suppose $y\operatorname{mod}p$\ and
$z\operatorname{mod}p$\ both assume at least $\left(  1+\gamma\right)
p/2$\ distinct values as we range over all $x\in\left\{  0,1\right\}  ^{n}$.
\ Then by the pigeonhole principle, for at least $\gamma p$\ possible values
of $y\operatorname{mod}p$, there exists a unique possible value of
$z\operatorname{mod}p$\ for which $y+z+c\equiv0\left(  \operatorname{mod}%
p\right)  $\ and hence $f_{R}\left(  y,z\right)  =1$. \ So
$\operatorname*{rank}\left(  M_{f|R}\right)  \geq\gamma p$, where $M_{f|R}%
$\ is the $2^{l}\times2^{l}$\ matrix whose $\left(  y,z\right)  $\ entry is
$f_{R}\left(  y,z\right)  $. \ It follows that assuming Conjecture
\ref{primes},%
\[
\Pr_{R\in\mathcal{R}_{l}}\left[  \operatorname*{rank}\left(  M_{f|R}\right)
\geq\gamma p\right]  =n^{-o\left(  \log n\right)  }.
\]
Furthermore, $\gamma p\geq2^{l-l^{1/8}/2}$\ for sufficiently large $n$ since
$p=\Omega\left(  2^{n^{\delta}}\right)  $. \ Therefore $\operatorname*{MFS}%
\left(  f\right)  =n^{\Omega\left(  \log n\right)  }$\ by Corollary
\ref{razcor2}.
\end{proof}

Using the ideas of Theorem \ref{ecclbapprox},\ one can show that under the
same conjecture, $\operatorname*{MFS}_{\varepsilon}\left(  f_{n,p}\right)
=n^{\Omega\left(  \log n\right)  }$\ and $\operatorname*{TS}_{\varepsilon
}\left(  \left\vert p\mathbb{Z}\right\rangle \right)  =n^{\Omega\left(  \log
n\right)  }$ for all $\varepsilon<1$---in other words, there exist Shor states
that cannot be approximated by polynomial-size trees.

In an earlier version of this paper, Conjecture \ref{primes}\ was stated
without any restriction on how the set $S$ is formed. \ The resulting
conjecture was far more general than we needed, and indeed was falsified by
Carl Pomerance (personal communication).

\subsection{Error Correction, Tree Size, and Persistence of
Entanglement\label{PERSIST}}

In this section we pursue a deeper understanding of our lower bounds. \ Recall
the states for which we were most successful in proving lower bounds are
exactly the states that arise in quantum error correction. \ Is this just a
coincidence, or should it have been expected? \ Also, can Raz's technique be
given any \textit{physical} interpretation?

Let $\left\vert S\right\rangle $\ be a uniform superposition over the elements
of some subset $S\subset\left\{  0,1\right\}  ^{n}$. \ Then our first
observation is that if the elements of $S$ are codewords of a sufficiently
good erasure code, then Corollary \ref{razcor2}\ yields an $n^{\Omega\left(
\log n\right)  }$\ tree size lower bound for $\left\vert S\right\rangle $.

\begin{theorem}
\label{eccgeneral}Let $l=n^{\delta}$\ for some $\delta\in\left(  0,\frac
{8}{15}\right)  $, and let $l<L<\frac{n}{4l^{7/8}}$. \ Suppose that
$\left\vert S\right\vert =2^{n-L}$\ (that is, $n-L$\ bits are being
encoded);\ and that\ for each $x\in S$, if we are given $n-l$\ bits of $x$
drawn uniformly at random together with their locations, then with probability
$1-o\left(  1\right)  $\ we can recover $x$ itself. \ Then $\operatorname*{TS}%
\left(  \left\vert S\right\rangle \right)  =n^{\Omega\left(  \log n\right)  }$.
\end{theorem}

\begin{proof}
Let $f\left(  x\right)  =1$\ if $x\in S$\ and $0$\ otherwise. \ Then it
suffices to show that \
\[
\Pr_{R\in\mathcal{R}_{l}}\left[  \operatorname*{rank}\left(  M_{f|R}\right)
\geq2^{l-l^{1/8}/2}\right]  =\Omega\left(  1\right)  .
\]
Clearly an $x\in S$\ drawn uniformly at random has entropy $n-L$. \ So if
$i_{1},\ldots,i_{n-l}\in\left\{  1,\ldots,n\right\}  $\ are drawn uniformly at
random without replacement, then the subsequence $x_{i_{1}},\ldots
,x_{i_{n-2l}}$\ has expected entropy $\frac{n-2l}{n}\left(  n-L\right)  $, and
the subsequence $x_{i_{1}},\ldots,x_{i_{n-l}}$\ has expected entropy
$\frac{n-l}{n}\left(  n-L\right)  $. \ By Markov's inequality, therefore, the
entropy of $x_{i_{n-2l+1}},\ldots,x_{i_{n-l}}$\ conditioned on $x_{i_{1}%
},\ldots,x_{i_{n-l}}$\ is at least $l-\frac{2lL}{n}$\ with probability at
least $1/2$ (since the entropy can never be greater than $l$). \ It follows
that with probability at least $1/2$\ over the restriction $R\in
\mathcal{R}_{l}$, there are at least $2^{l-2lL/n}>2^{l-l^{1/8}/2}$\ distinct
settings of $y\in\left\{  0,1\right\}  ^{l}$\ for which $f_{R}\left(
y,z\right)  =1$.\ \ Here we have used the fact that $L<\frac{n}{4l^{7/8}}$.
\ But this then implies that $\operatorname*{rank}\left(  M_{f|R}\right)
\geq2^{l-l^{1/8}/2}$\ with probability $1-o\left(  1\right)  $. \ For given
$y$, if there\ are two or more values of $z$ for which $f_{R}\left(
y,z\right)  =1$, then $x$ is not uniquely recoverable from the $n-l$\ bits
outside of $z$.
\end{proof}

The converse of Theorem \ref{eccgeneral}\ is false. \ For choose
$S\subset\left\{  0,1\right\}  ^{n}$\ uniformly at random subject to
$\left\vert S\right\vert =2^{n-1}$. \ Then Corollary \ref{razcor2}\ yields an
$n^{\Omega\left(  \log n\right)  }$\ lower bound on $\operatorname*{TS}\left(
\left\vert S\right\rangle \right)  $, but $S$ does not correspond to any good
error-correcting code. \ So roughly speaking, if $\left\vert S\right\rangle
$\ is a codeword state then $\left\vert S\right\rangle $\ has large tree size,
but not vice versa.

We can gain further insight by asking what \textit{physical} properties a
codeword state has to have. \ One important property is \textquotedblleft
persistence of entanglement,\textquotedblright\ introduced D\"{u}r and Briegel
\cite{db} among others. \ This is the property of remaining highly entangled
even after a limited amount of interaction with the environment. \ For
example, the Schr\"{o}dinger cat state $\left(  \left\vert 0\right\rangle
^{\otimes n}+\left\vert 1\right\rangle ^{\otimes n}\right)  /\sqrt{2}$\ is in
some sense highly entangled, but it is \textit{not} persistently entangled,
since measuring a single qubit in the standard basis destroys all entanglement.

By contrast, consider the \textquotedblleft cluster states\textquotedblright%
\ defined by Briegel and Raussendorf \cite{br}. \ These states have attracted
a great deal of attention because of their application to quantum computing
via $1$-qubit measurements only \cite{rbb}. \ For our purposes, a
two-dimensional cluster state is an equal superposition over all settings of a
$\sqrt{n}\times\sqrt{n}$\ array of bits,\ with each basis state having a phase
of $\left(  -1\right)  ^{r}$, where $r$ is the number of horizontally or
vertically adjacent pairs of bits that are both `$1$'. \ D\"{u}r and Briegel
\cite{db} showed that such states are persistently entangled in a precise
sense:\ one can distill $n$-partite entanglement from them even after each
qubit has interacted with a heat bath for an amount of time independent of $n$.

Persistence of entanglement seems related to how one shows tree size lower
bounds using Raz's technique. \ For to apply Corollary \ref{razcor2}, one
basically \textquotedblleft measures\textquotedblright\ most of a state's
qubits, then partitions the unmeasured qubits into two subsystems of equal
size, and argues that with high probability those two subsystems are still
almost maximally entangled. \ The connection is not perfect, though. \ For one
thing, setting most of the qubits to $0$ or $1$ uniformly at random is not the
same as measuring them. \ For another, Theorem \ref{razthm}\ yields
$n^{\Omega\left(  \log n\right)  }$\ tree size lower bounds without the need
to trace out a subset of qubits. \ It suffices for the \textit{original} state
to be almost maximally entangled, no matter how one partitions it into two
subsystems of equal size.

But what about $2$-D cluster states---do \textit{they} have tree size
$n^{\Omega\left(  \log n\right)  }$? \ We strongly conjecture that the answer
is `yes.' \ However, proving this conjecture will almost certainly require
going beyond Theorem \ref{razthm}. \ One will want to use random restrictions
that respect the $2$-D neighborhood structure of cluster states---similar to
the restrictions used by Raz \cite{raz}\ to show that permanent and
determinant have multilinear formula size $n^{\Omega\left(  \log n\right)  }$.

We end this section by showing that there exist states that are persistently
entangled in the sense of D\"{u}r and Briegel \cite{db}, but that have
polynomial tree size. \ In particular, D\"{u}r and Briegel showed that even
\textit{one}-dimensional cluster states are persistently entangled. \ On the
other hand:

\begin{proposition}
\label{1d}Let%
\[
\left\vert \psi\right\rangle =\frac{1}{2^{n/2}}\sum_{x\in\left\{  0,1\right\}
^{n}}\left(  -1\right)  ^{x_{1}x_{2}+x_{2}x_{3}+\cdots+x_{n-1}x_{n}}\left\vert
x\right\rangle .
\]
Then $\operatorname*{TS}\left(  \left\vert \psi\right\rangle \right)
=O\left(  n^{4}\right)  $.
\end{proposition}

\begin{proof}
Given bits $i,j,k$, let $\left\vert P_{n}^{ijk}\right\rangle $\ be an equal
superposition over all $n$-bit strings $x_{1}\ldots x_{n}$\ such that
$x_{1}=i$, $x_{n}=k$, and $x_{1}x_{2}+\cdots+x_{n-1}x_{n}\equiv j\left(
\operatorname{mod}2\right)  $. \ Then%
\begin{align*}
\left\vert P_{n}^{i0k}\right\rangle  &  =\frac{1}{\sqrt{8}}\left(
\begin{array}
[c]{c}%
\left\vert P_{n/2}^{i00}\right\rangle \left\vert P_{n/2}^{00k}\right\rangle
+\left\vert P_{n/2}^{i10}\right\rangle \left\vert P_{n/2}^{01k}\right\rangle
+\left\vert P_{n/2}^{i00}\right\rangle \left\vert P_{n/2}^{10k}\right\rangle
+\left\vert P_{n/2}^{i10}\right\rangle \left\vert P_{n/2}^{11k}\right\rangle
+\\
\left\vert P_{n/2}^{i01}\right\rangle \left\vert P_{n/2}^{00k}\right\rangle
+\left\vert P_{n/2}^{i11}\right\rangle \left\vert P_{n/2}^{01k}\right\rangle
+\left\vert P_{n/2}^{i01}\right\rangle \left\vert P_{n/2}^{11k}\right\rangle
+\left\vert P_{n/2}^{i11}\right\rangle \left\vert P_{n/2}^{10k}\right\rangle
\end{array}
\right)  ,\\
\left\vert P_{n}^{i1k}\right\rangle  &  =\frac{1}{\sqrt{8}}\left(
\begin{array}
[c]{c}%
\left\vert P_{n/2}^{i00}\right\rangle \left\vert P_{n/2}^{01k}\right\rangle
+\left\vert P_{n/2}^{i10}\right\rangle \left\vert P_{n/2}^{00k}\right\rangle
+\left\vert P_{n/2}^{i00}\right\rangle \left\vert P_{n/2}^{11k}\right\rangle
+\left\vert P_{n/2}^{i10}\right\rangle \left\vert P_{n/2}^{10k}\right\rangle
+\\
\left\vert P_{n/2}^{i01}\right\rangle \left\vert P_{n/2}^{01k}\right\rangle
+\left\vert P_{n/2}^{i11}\right\rangle \left\vert P_{n/2}^{00k}\right\rangle
+\left\vert P_{n/2}^{i01}\right\rangle \left\vert P_{n/2}^{10k}\right\rangle
+\left\vert P_{n/2}^{i11}\right\rangle \left\vert P_{n/2}^{11k}\right\rangle
\end{array}
\right)  .
\end{align*}
Therefore $\operatorname*{TS}\left(  \left\vert P_{n}^{ijk}\right\rangle
\right)  \leq16\operatorname*{TS}\left(  \left\vert P_{n/2}^{ijk}\right\rangle
\right)  $, and solving this recurrence relation yields $\operatorname*{TS}%
\left(  \left\vert P_{n}^{ijk}\right\rangle \right)  =O\left(  n^{4}\right)
$. \ Finally observe that%
\[
\left\vert \psi\right\rangle =\left(  \frac{\left\vert 0\right\rangle
+\left\vert 1\right\rangle }{\sqrt{2}}\right)  ^{\otimes n}-\frac{\left\vert
P_{n}^{010}\right\rangle +\left\vert P_{n}^{011}\right\rangle +\left\vert
P_{n}^{110}\right\rangle +\left\vert P_{n}^{111}\right\rangle }{\sqrt{2}}.
\]

\end{proof}

\section{Computing With Tree States\label{TREEBQP}}

Suppose a quantum computer is restricted to being in a tree state at all
times. \ (We can imagine that if the tree size ever exceeds some polynomial
bound, the quantum computer explodes, destroying our laboratory.) \ Does the
computer then have an efficient classical simulation? \ In other words,
letting $\mathsf{TreeBQP}$\ be the class of languages accepted by such a
machine, does $\mathsf{TreeBQP=BPP}$? \ A positive answer would make tree
states more attractive as a Sure/Shor separator. \ For once we admit any
states incompatible with the polynomial-time Church-Turing thesis, it seems
like we might as well go all the way, and admit \textit{all} states preparable
by polynomial-size quantum circuits! \ The $\mathsf{TreeBQP}$\ versus
$\mathsf{BPP}$\ problem is closely related to the problem of finding an
efficient (classical) algorithm to \textit{learn} multilinear formulas. \ In
light of Raz's lower bound, and of the connection between lower bounds and
learning noticed by Linial, Mansour, and Nisan \cite{lmn}, the latter problem
might be less hopeless than it looks. \ In this section we show a weaker
result: that $\mathsf{TreeBQP}$\ is contained in $\mathsf{\Sigma}%
_{3}^{\mathsf{P}}\cap\mathsf{\Pi}_{3}^{\mathsf{P}}$, the third level of the
polynomial hierarchy. \ Since $\mathsf{BQP}$\ is not known to lie in
$\mathsf{PH}$, this result could be taken as weak evidence that
$\mathsf{TreeBQP\neq BQP}$. \ (On the other hand, we do not yet have oracle
evidence even for\ $\mathsf{BQP}\not \subset \mathsf{AM}$, though not for lack
of trying \cite{aarrfs}.)

\begin{definition}
\label{formulabqp}$\mathsf{TreeBQP}$\ is the class of languages accepted by a
$\mathsf{BQP}$\ machine subject to the constraint that at every time step $t$,
the machine's state $\left\vert \psi^{\left(  t\right)  }\right\rangle $\ is
exponentially close to a tree state. \ More formally, the initial state is
$\left\vert \psi^{\left(  0\right)  }\right\rangle =\left\vert 0\right\rangle
^{\otimes\left(  p\left(  n\right)  -n\right)  }\otimes\left\vert
x\right\rangle $ (for an input $x\in\left\{  0,1\right\}  ^{n}$\ and
polynomial bound $p$), and a uniform classical polynomial-time algorithm
generates a sequence of gates $g^{\left(  1\right)  },\ldots,g^{\left(
p\left(  n\right)  \right)  }$. \ Each $g^{\left(  t\right)  }$\ can be either
be selected from some finite universal basis of unitary gates (as we will show
in Theorem \ref{nice}, part (i), the choice of gate set will not matter), or
can be a $1$-qubit measurement. \ When we perform a measurement, the state
evolves to one of two possible pure states, with the usual probabilities,
rather than to a mixed state. \ We require that the final gate $g^{\left(
p\left(  n\right)  \right)  }$\ is a measurement of the first qubit. \ If at
least one intermediate state $\left\vert \psi^{\left(  t\right)
}\right\rangle $\ had\ $\operatorname*{TS}_{1/2^{\Omega\left(  n\right)  }%
}\left(  \left\vert \psi^{\left(  t\right)  }\right\rangle \right)  >p\left(
n\right)  $,\ then the outcome of the final measurement is chosen
adversarially; otherwise it is given by the usual Born probabilities. \ The
measurement must return $1$\ with probability at least $2/3$\ if the input is
in the language, and with probability at most $1/3$\ otherwise.
\end{definition}

Some comments on the definition: we allow $\left\vert \psi^{\left(  t\right)
}\right\rangle $\ to deviate from a tree state by an exponentially small
amount, in order to make the model independent of the choice of gate set. \ We
allow intermediate measurements because otherwise it is unclear even how to
simulate $\mathsf{BPP}$.\footnote{If we try to simulate $\mathsf{BPP}$ in the
standard way, we might produce complicated entanglement between the
computation register and the register containing the random bits, and no
longer have a tree state.} \ The rule for measurements follows the
\textquotedblleft Copenhagen interpretation,\textquotedblright\ in the sense
that if a qubit is measured to be $1$, then subsequent computation is not
affected by what would have happened were the qubit measured to be $0$. \ In
particular, if measuring $0$ would have led to states of tree size greater
than $p\left(  n\right)  $, that does not invalidate the results of the path
where $1$ is measured.

The following theorem shows that $\mathsf{TreeBQP}$\ has many of the
properties we would want it to have.

\begin{theorem}
\label{nice}\quad

\begin{enumerate}
\item[(i)] The definition of $\mathsf{TreeBQP}$\ is invariant under the choice
of gate set.

\item[(ii)] The probabilities $\left(  1/3,2/3\right)  $\ can be replaced by
any $\left(  p,1-p\right)  $ with $2^{-2^{\sqrt{\log n}}}<p<1/2$.

\item[(iii)] $\mathsf{BPP}\subseteq\mathsf{TreeBQP}\subseteq\mathsf{BQP}$.
\end{enumerate}
\end{theorem}

\begin{proof}
\quad

\begin{enumerate}
\item[(i)] The Solovay-Kitaev Theorem \cite{kitaev,nc} shows that given a
universal gate set, we can approximate any $k$-qubit unitary to accuracy
$1/\varepsilon$\ using $k$\ qubits and a circuit of size $O\left(
\operatorname*{polylog}\left(  1/\varepsilon\right)  \right)  $. \ So let
$\left\vert \psi^{\left(  0\right)  }\right\rangle ,\ldots,\left\vert
\psi^{\left(  p\left(  n\right)  \right)  }\right\rangle \in\mathcal{H}%
_{2}^{\otimes p\left(  n\right)  }$ be a sequence of states, with $\left\vert
\psi^{\left(  t\right)  }\right\rangle $\ produced from $\left\vert
\psi^{\left(  t-1\right)  }\right\rangle $\ by applying a $k$-qubit unitary
$g^{\left(  t\right)  }$\ (where $k=O\left(  1\right)  $).\ \ Then using a
polynomial-size circuit, we can approximate each $\left\vert \psi^{\left(
t\right)  }\right\rangle $ to accuracy $1/2^{\Omega\left(  n\right)  }$, as in
the definition of $\mathsf{TreeBQP}$. \ Furthermore, since the approximation
circuit for $g^{\left(  t\right)  }$ acts only on $k$\ qubits, any
intermediate state $\left\vert \varphi\right\rangle $\ it produces satisfies
$\operatorname*{TS}\nolimits_{1/2^{\Omega\left(  n\right)  }}\left(
\left\vert \varphi\right\rangle \right)  \leq k4^{k}\operatorname*{TS}%
\nolimits_{1/2^{\Omega\left(  n\right)  }}\left(  \left\vert \psi^{\left(
t-1\right)  }\right\rangle \right)  $\ by Proposition \ref{invariant}.

\item[(ii)] To amplify to a constant probability, run $k$ copies of the
computation in tensor product, then output the majority answer. \ By part (i),
outputting the majority can increase the tree size by a factor of at
most\ $2^{k+1}$. \ To amplify to $2^{-2^{\sqrt{\log n}}}$, observe that the
Boolean majority function on $k$ bits has a multilinear formula of size
$k^{O\left(  \log k\right)  }$. \ For let $T_{k}^{h}\left(  x_{1},\ldots
,x_{k}\right)  $\ equal $1$ if $x_{1}+\cdots+x_{k}\geq h$\ and $0$ otherwise;
then%
\[
T_{k}^{h}\left(  x_{1},\ldots,x_{k}\right)  =1-\prod_{i=0}^{h}\left(
1-T_{\left\lfloor k/2\right\rfloor }^{i}\left(  x_{1},\ldots,x_{\left\lfloor
k/2\right\rfloor }\right)  T_{\left\lceil k/2\right\rceil }^{h-i}\left(
x_{\left\lfloor k/2\right\rfloor +1},\ldots,x_{k}\right)  \right)  ,
\]
so $\operatorname*{MFS}\left(  T_{k}^{h}\right)  \leq2h\max_{i}%
\operatorname*{MFS}\left(  T_{\left\lceil k/2\right\rceil }^{h}\right)
+O\left(  1\right)  $, and solving this recurrence yields $\operatorname*{MFS}%
\left(  T_{k}^{k/2}\right)  =k^{O\left(  \log k\right)  }$. \ Substituting
$k=2^{\sqrt{\log n}}$\ into $k^{O\left(  \log k\right)  }$\ yields
$n^{O\left(  1\right)  }$, meaning the tree size increases by at most a
polynomial factor.

\item[(iii)] To simulate $\mathsf{BPP}$, we just perform a classical
reversible computation, applying a Hadamard followed by a measurement to some
qubit whenever we need a random bit. \ Since the number of basis states with
nonzero amplitude is at most $2$, the simulation is clearly in
$\mathsf{TreeBQP}$. \ The other containment is obvious.
\end{enumerate}
\end{proof}

\begin{theorem}
\label{inph}$\mathsf{TreeBQP}\subseteq\mathsf{\Sigma}_{3}^{\mathsf{P}}%
\cap\mathsf{\Pi}_{3}^{\mathsf{P}}$.
\end{theorem}

\begin{proof}
Since $\mathsf{TreeBQP}$\ is closed under complement, it suffices to show that
$\mathsf{TreeBQP}\subseteq\mathsf{\Pi}_{3}^{\mathsf{P}}$. \ Our proof will
combine approximate counting with a predicate to verify the correctness of a
$\mathsf{TreeBQP}$\ computation. \ Let $C$ be a uniformly-generated quantum
circuit, and let $M=\left(  m^{\left(  1\right)  },\ldots,m^{\left(  p\left(
n\right)  \right)  }\right)  $\ be a sequence of binary measurement outcomes.
\ We adopt the convention that after making a measurement, the state vector is
\textit{not} rescaled to have norm $1$. \ That way the probabilities across
all `measurement branches' continue to sum to $1$. \ Let $\left\vert
\psi_{M,x}^{\left(  0\right)  }\right\rangle ,\ldots,\left\vert \psi
_{M,x}^{\left(  p\left(  n\right)  \right)  }\right\rangle $\ be the sequence
of unnormalized pure states under measurement outcome sequence $M$ and input
$x$, where $\left\vert \psi_{M,x}^{\left(  t\right)  }\right\rangle
=\sum_{y\in\left\{  0,1\right\}  ^{p\left(  n\right)  }}\alpha_{y,M,x}%
^{\left(  t\right)  }\left\vert y\right\rangle $. \ Also, let $\Lambda\left(
M,x\right)  $\ express that $\operatorname*{TS}_{1/2^{\Omega\left(  n\right)
}}\left(  \left\vert \psi_{M,x}^{\left(  t\right)  }\right\rangle \right)
\leq p\left(  n\right)  $ for every $t$. \ Then $C$ accepts if%
\[
W_{x}=\sum_{M\,:\,\Lambda\left(  M,x\right)  }\sum_{y\in\left\{  0,1\right\}
^{p\left(  n\right)  -1}}\left\vert \alpha_{1y,M,x}^{\left(  p\left(
n\right)  \right)  }\right\vert ^{2}\geq\frac{2}{3},
\]
while $C$ rejects if $W_{x}\leq1/3$. \ If we could compute each $\left\vert
\alpha_{1y,M,x}^{\left(  p\left(  n\right)  \right)  }\right\vert $
efficiently (as well as $\Lambda\left(  M,x\right)  $), we would then have a
$\mathsf{\Pi}_{2}^{\mathsf{P}}$\ predicate expressing that $W_{x}\geq2/3$.
\ This follows since we can do approximate counting via hashing in
$\mathsf{AM}\subseteq\mathsf{\Pi}_{2}^{\mathsf{P}}$\ \cite{gs}, and thereby
verify that an exponentially large sum of nonnegative terms is at least $2/3$,
rather than at most $1/3$. \ The one further fact we need is that in our
$\mathsf{\Pi}_{2}^{\mathsf{P}}$\ ($\forall\exists$) predicate, we can take the
existential quantifier to range over tuples of `candidate solutions'---that
is, $\left(  M,y\right)  $\ pairs together with lower bounds $\beta$ on
$\left\vert \alpha_{1y,M,x}^{\left(  p\left(  n\right)  \right)  }\right\vert
$.

It remains only to show how we verify that $\Lambda\left(  M,x\right)
$\ holds and that $\left|  \alpha_{1y,M,x}^{\left(  p\left(  n\right)
\right)  }\right|  =\beta$. \ First, we extend the existential quantifier so
that it guesses not only $M$ and $y$, but also a sequence of trees $T^{\left(
0\right)  },\ldots,T^{\left(  p\left(  n\right)  \right)  }$, representing
$\left|  \psi_{M,x}^{\left(  0\right)  }\right\rangle ,\ldots,\left|
\psi_{M,x}^{\left(  p\left(  n\right)  \right)  }\right\rangle $ respectively.
\ Second, using the last universal quantifier to range over $\widehat{y}%
\in\left\{  0,1\right\}  ^{p\left(  n\right)  }$, we verify the following:

\begin{enumerate}
\item[(1)] $T^{\left(  0\right)  }$ is a fixed tree representing $\left|
0\right\rangle ^{\otimes\left(  p\left(  n\right)  -n\right)  }\otimes\left|
x\right\rangle $.

\item[(2)] $\left|  \alpha_{1y,M,x}^{\left(  p\left(  n\right)  \right)
}\right|  $ equals its claimed value to $\Omega\left(  n\right)  $\ bits of precision.

\item[(3)] Let $g^{\left(  1\right)  },\ldots,g^{\left(  p\left(  n\right)
\right)  }$\ be the gates applied by $C$. \ Then for all $t$ and $\widehat{y}%
$, if $g^{\left(  t\right)  }$\ is unitary then $\alpha_{\widehat{y}%
,M,x}^{\left(  t\right)  }=\left\langle \widehat{y}\right\vert \cdot
g^{\left(  t\right)  }\left\vert \psi_{M,x}^{\left(  t-1\right)
}\right\rangle $ to $\Omega\left(  n\right)  $\ bits of precision. \ Here the
right-hand side is a sum of $2^{k}$\ terms ($k$ being the number of qubits
acted on by $g^{\left(  t\right)  }$), each term efficiently computable given
$T^{\left(  t-1\right)  }$. \ Similarly, if $g^{\left(  t\right)  }$\ is a
measurement of the $i^{th}$\ qubit, then $\alpha_{\widehat{y},M,x}^{\left(
t\right)  }=\alpha_{\widehat{y},M,x}^{\left(  t-1\right)  }$\ if the $i^{th}%
$\ bit of $\widehat{y}$\ equals $m^{\left(  t\right)  }$, while $\alpha
_{\widehat{y},M,x}^{\left(  t\right)  }=0$\ otherwise.
\end{enumerate}
\end{proof}

In the proof of Theorem \ref{inph}, the only fact about tree states we use is
that $\mathsf{Tree}\subseteq\mathsf{AmpP}$; that is, there is a
polynomial-time classical algorithm that computes the amplitude $\alpha_{x}%
$\ of any basis state $\left\vert x\right\rangle $. \ So if we define
$\mathsf{AmpP}$-$\mathsf{BQP}$\ analogously to $\mathsf{TreeBQP}$ except that
any states in $\mathsf{AmpP}$\ are allowed, then $\mathsf{AmpP}$%
-$\mathsf{BQP}\subseteq\mathsf{\Sigma}_{3}^{\mathsf{P}}\cap\mathsf{\Pi}%
_{3}^{\mathsf{P}}$\ as well.

\section{The Experimental Situation\label{EXPER}}

The results of this paper suggest an obvious challenge for experimenters:
\textit{prepare non-tree states in the lab}. \ For were this challenge met, it
would rule out one way in which quantum mechanics could fail, just as the Bell
inequality experiments of Aspect et al. \cite{aspect}\ did twenty years ago.
\ If they wished, quantum computing skeptics could then propose a new
candidate Sure/Shor separator, and experimenters could try to rule out
\textit{that} one, and so on. \ The result would be to divide the question of
whether quantum computing is possible into a series of smaller questions about
which states can be prepared. \ In our view, this would aid progress in two
ways: by helping experimenters set clear goals, and by forcing theorists to
state clear positions.

However, our experimental challenge raises some immediate questions.\ \ In
particular, what would it \textit{mean} to prepare a non-tree state? \ How
would we know if we succeeded? \ Also, have non-tree states already been
prepared (or observed)? \ The purpose of this section is to set out our
thoughts about these questions.

First of all, when discussing experiments, it goes without saying that we must
convert asymptotic statements into statements about specific values of $n$.
\ The central tenet of computational complexity theory is that this is
possible. \ Thus, instead of asking whether $n$-qubit states with tree size
$2^{\Omega\left(  n\right)  }$\ can be prepared, we ask whether $200$-qubit
states with tree size at least (say) $2^{80}$\ can be prepared. \ Even though
the second question does not logically imply anything about the first, the
second is closer to what we ultimately care about anyway. \ Admittedly,
knowing that $\operatorname*{TS}\left(  \left\vert \psi_{n}\right\rangle
\right)  =n^{\Omega\left(  \log n\right)  }$\ tells us little about
$\operatorname*{TS}\left(  \left\vert \psi_{100}\right\rangle \right)  $\ or
$\operatorname*{TS}\left(  \left\vert \psi_{200}\right\rangle \right)  $,
especially since in Raz's paper \cite{raz}, the constant in the exponent
$\Omega\left(  \log n\right)  $\ is taken to be $10^{-6}$\ (though this can
certainly be improved). \ Thus, proving tight lower bounds for small $n$ is
one of the most important problems left open by this paper. \ In Appendix
\ref{MOTS} we solve the problem for the case of manifestly orthogonal tree size.

A second common objection is that our formalism applies only to pure states,
but in reality all states are mixed. \ However, there are several natural ways
to extend the formalism to mixed states. \ \ Given a mixed state $\rho$, we
could minimize tree size over all purifications of $\rho$, or minimize the
expected tree size $\sum_{i}\left\vert \alpha_{i}\right\vert ^{2}%
\operatorname*{TS}\left(  \left\vert \psi_{i}\right\rangle \right)  $, or
maximum $\max_{i}\operatorname*{TS}\left(  \left\vert \psi_{i}\right\rangle
\right)  $, over all decompositions $\rho=\sum_{i}\alpha_{i}\left\vert
\psi_{i}\right\rangle \left\langle \psi_{i}\right\vert $.

A third objection is a real quantum state might be a \textquotedblleft
soup\textquotedblright\ of free-wandering fermions and bosons, with no
localized subsystems corresponding to qubits. \ How can one determine the tree
size of such a state? \ The answer is that one cannot. \ Any complexity
measure for particle position and momentum states would have to be quite
different from the measures considered in this paper. \ On the other hand, the
states of interest for quantum computing usually \textit{do} involve localized
qubits. \ Indeed, even if quantum information is stored in particle positions,
one might force each particle into two sites (corresponding to $\left\vert
0\right\rangle $\ and $\left\vert 1\right\rangle $), neither of which can be
occupied by any other particle. \ In that case it again becomes meaningful to
discuss tree size.

But how do we verify that a state with large tree size was prepared? \ Of
course, if $\left\vert \psi\right\rangle $\ is preparable by a polynomial-size
quantum circuit, then \textit{assuming quantum mechanics is valid} (and
assuming our gates behave as specified), we can always test whether a given
state $\left\vert \varphi\right\rangle $\ is close to $\left\vert
\psi\right\rangle $\ or not. \ Let $U$\ map $\left\vert 0\right\rangle
^{\otimes n}$\ to $\left\vert \psi\right\rangle $; then it suffices to test
whether $U^{-1}\left\vert \varphi\right\rangle $\ is close to $\left\vert
0\right\rangle ^{\otimes n}$. \ However, in the experiments under discussion,
the validity of quantum mechanics is the very point in question. \ And once we
allow Nature to behave in arbitrary ways, a skeptic could explain \textit{any}
experimental result without having to invoke states with large tree size.

The above fact has often been urged against us, but as it stands, it is no
different from the fact that one could explain any astronomical observation
without abandoning the Ptolemaic system. \ The issue is not one of
mathematical proof, but of accumulating observations that are consistent
with\ the hypothesis of large tree size, and inconsistent with alternative
hypotheses if we disallow special pleading. \ So for example, to test whether
the subgroup state%
\[
\left\vert S\right\rangle =\frac{1}{\sqrt{\left\vert S\right\vert }}\sum_{x\in
S}\left\vert x\right\rangle
\]
was prepared, we might use CNOT gates to map $\left\vert x\right\rangle $\ to
$\left\vert x\right\rangle \left\vert v^{T}x\right\rangle $ for some vector
$v\in\mathbb{Z}_{2}^{n}$. \ Based on our knowledge of $S$, we could then
predict whether the qubit $\left\vert v^{T}x\right\rangle $\ should be
$\left\vert 0\right\rangle $, $\left\vert 1\right\rangle $, or an equal
mixture of $\left\vert 0\right\rangle $ and $\left\vert 1\right\rangle $ when
measured. \ Or we could apply Hadamard gates to all $n$ qubits of $\left\vert
S\right\rangle $, then perform the same test for the subgroup dual to $S$.
\ In saying that a system is in state $\left\vert S\right\rangle $, it is not
clear if we \textit{mean} anything more than that it responds to all such
tests in expected ways. \ Similar remarks apply to Shor states and cluster states.

In our view, tests of the sort described above are certainly
\textit{sufficient}, so the interesting question is whether they are
\textit{necessary}, or whether weaker and more indirect tests would also
suffice. \ This question rears its head when we ask whether non-tree states
have already been observed. \ For as pointed out to us by Anthony Leggett,
there exist systems studied in condensed-matter physics that are strong
candidates for having superpolynomial tree size. \ An example is the magnetic
salt LiHo$_{x}$Y$_{1-x}$F$_{4}$ studied by Ghosh et al. \cite{grac}, which,
like the cluster states of Briegel and Raussendorf \cite{br}, basically
consists of a lattice of spins subject to pairwise nearest-neighbor
Hamiltonians. \ The main differences are that the salt lattice is 3-D instead
of 2-D, is tetragonal instead of cubic, and is irregular in that not every
site is occupied by a spin. \ Also, there are weak interactions even between
spins that are not nearest neighbors. \ But none of these differences seems
likely to change a superpolynomial tree size into a polynomial one.

For us, the main issues are (1) how precisely can we characterize\footnote{By
\textquotedblleft characterize,\textquotedblright\ we mean give an explicit
formula for the amplitudes at a particular time $t$, in some standard basis.
\ If a state is characterized as the ground state of a Hamiltonian, then we
first need to solve for the amplitudes before we can prove tree size lower
bounds using Raz's method.} the quantum state of the magnetic salt, and (2)
how strong the evidence is that that \textit{is} the state. \ What Ghosh et
al. \cite{grac}\ did was to calculate bulk properties of the salt, such as its
magnetic susceptibility and specific heat, with and without taking into
account the quantum entanglement generated by the nearest-neighbor
Hamiltonians. \ They found that including entanglement yielded a better fit to
the experimentally measured values. \ However, this is clearly a far cry from
preparing a system in a state of one's choosing by applying a known pulse
sequence, and then applying any of a vast catalog of tests to verify that the
state was prepared. \ So it would be valuable to have more direct evidence
that states qualitatively like cluster states can exist in Nature.

In summary, our results underscore the importance of current experimental work
on large, persistently entangled quantum states; but they also suggest a new
motivation and perspective for this work. \ They suggest that we reexamine
known condensed-matter systems with a new goal in mind: understanding the
complexity of their associated quantum states. \ They also suggest that 2-D
cluster states and random subgroup states are interesting in a way that 1-D
spin chains and Schr\"{o}dinger cat states are not. \ Yet when experimenters
try to prepare states of the former type, they often see it as merely a
stepping stone towards demonstrating error-correction or another quantum
computing benchmark. \ Thus, Knill et al. \cite{klmn}
prepared\footnote{Admittedly, what they really prepared is the `pseudo-pure'
state $\rho=\varepsilon\left\vert \psi\right\rangle \left\langle
\psi\right\vert +\left(  1-\varepsilon\right)  I$, where $I$ is the maximally
mixed state and $\varepsilon\approx10^{-5}$. \ Braunstein et al.
\cite{bcjlps}\ have shown that, if the number of qubits $n$ is less than about
$14$, then such states cannot be entangled. \ That is, there exists a
representation of $\rho$\ as a mixture of pure states, each of which is
separable and therefore has tree size $O\left(  n\right)  $. \ This is a
well-known limitation of the liquid NMR technology used by Knill et al.
\ Thus, a key challenge is to replicate the successes of liquid NMR using
colder qubits.} the $5$-qubit state%
\[
\left\vert \psi\right\rangle =\frac{1}{4}\left(
\begin{array}
[c]{c}%
\left\vert 00000\right\rangle +\left\vert 10010\right\rangle +\left\vert
01001\right\rangle +\left\vert 10100\right\rangle +\left\vert
01010\right\rangle -\left\vert 11011\right\rangle -\left\vert
00110\right\rangle -\left\vert 11000\right\rangle \\
-\left\vert 11101\right\rangle -\left\vert 00011\right\rangle -\left\vert
11110\right\rangle -\left\vert 01111\right\rangle -\left\vert
10001\right\rangle -\left\vert 01100\right\rangle -\left\vert
10111\right\rangle +\left\vert 00101\right\rangle
\end{array}
\right)  ,
\]
for which $\operatorname*{MOTS}\left(  \left\vert \psi\right\rangle \right)
=40$\ from the decomposition%
\[
\left\vert \psi\right\rangle =\frac{1}{4}\left(
\begin{array}
[c]{c}%
\left(  \left\vert 01\right\rangle +\left\vert 10\right\rangle \right)
\otimes\left(  \left\vert 010\right\rangle -\left\vert 111\right\rangle
\right)  +\left(  \left\vert 01\right\rangle -\left\vert 10\right\rangle
\right)  \otimes\left(  \left\vert 001\right\rangle -\left\vert
100\right\rangle \right)  \\
-\left(  \left\vert 00\right\rangle +\left\vert 11\right\rangle \right)
\otimes\left(  \left\vert 011\right\rangle +\left\vert 110\right\rangle
\right)  +\left(  \left\vert 00\right\rangle -\left\vert 11\right\rangle
\right)  \otimes\left(  \left\vert 000\right\rangle +\left\vert
101\right\rangle \right)
\end{array}
\right)  ,
\]
and for which we conjecture $\operatorname*{TS}\left(  \left\vert
\psi\right\rangle \right)  =40$ as well. \ However, the sole motivation of the
experiment was to demonstrate a $5$-qubit quantum error-correcting code. \ In
our opinion, whether states with large tree size can be prepared is a
fundamental question in its own right. \ Were that question studied directly,
perhaps we could address it for larger numbers of qubits.

Let us end by stressing that, in the perspective we are advocating, there is
nothing sacrosanct about tree size as opposed to other complexity measures.
\ This paper concentrated on tree size because it is the subject of our main
results, and because it is better to be specific than vague. \ On the other
hand, Section \ref{BASIC},\ Appendix \ref{REL}, and Appendix \ref{MOTS}%
\ contain numerous results about orthogonal tree size, manifestly orthogonal
tree size, Vidal's $\chi$\ complexity, and other measures. \ Readers
dissatisfied with \textit{all} of these measures are urged to propose new
ones, perhaps motivated directly by experiments. \ We see nothing wrong with
having multiple ways to quantify the complexity of quantum states, and much
wrong with having no ways.

\section{Conclusion and Open Problems\label{OPEN}}

A crucial step in quantum computing was to separate the question of whether
quantum computers can be built from the question of what one could do with
them. \ This separation allowed computer scientists to make great advances on
the latter question, despite knowing nothing about the former. \ We have
argued, however, that the tools of computational complexity theory are
relevant to both questions. \ The claim that large-scale quantum computing is
possible in principle is really a claim that certain \textit{states} can
exist---that quantum mechanics will not break down if we try to prepare those
states. \ Furthermore, what distinguishes these states from states we have
seen must be more than precision in amplitudes, or the number of qubits
maintained coherently. \ The distinguishing property should instead be some
sort of \textit{complexity}. \ That is, Sure states should have succinct
representations of a type that Shor states do not.

We have tried to show that, by adopting this viewpoint, we make the debate
about whether quantum computing is possible less ideological and more
scientific. \ By studying particular examples of Sure/Shor separators, quantum
computing skeptics would strengthen their case---for they would then have a
plausible research program aimed at identifying what, exactly, the barriers to
quantum computation are. \ We hope, however, that the `complexity theory of
quantum states' initiated in this paper will be taken up by quantum computing
proponents as well. \ This theory offers a new perspective on the transition
from classical to quantum computing, and a new connection between quantum
computing and the powerful circuit lower bound techniques of classical
complexity theory.

We end with some open problems.

\begin{enumerate}
\item[(1)] Can Raz's technique be improved to show exponential tree size lower bounds?

\item[(2)] Can we prove Conjecture \ref{primes}, implying an $n^{\Omega\left(
\log n\right)  }$\ tree size lower bound for Shor states?

\item[(3)] Let $\left\vert \varphi\right\rangle $\ be a uniform superposition
over all $n$-bit strings of Hamming weight $n/2$. \ It is easy to show by
divide-and-conquer that $\operatorname*{TS}\left(  \left\vert \varphi
\right\rangle \right)  =n^{O\left(  \log n\right)  }$. \ Is this upper bound
tight? \ More generally, can we show a superpolynomial tree size lower bound
for any state with permutation symmetry?

\item[(4)] Is $\mathsf{Tree}=\mathsf{OTree}$? \ That is, are there tree states
that are not orthogonal tree states?

\item[(5)] Is the tensor-sum hierarchy of Section \ref{CQS}\ infinite? \ That
is, do we have $\mathsf{\Sigma}${}$_{\mathsf{k}}\neq\mathsf{\Sigma}$%
{}$_{\mathsf{k+1}}$\ for all $k$?

\item[(6)] Is $\mathsf{TreeBQP}=\mathsf{BPP}$? \ That is, can a quantum
computer that is always in a tree state be simulated classically? \ The key
question seems to be whether the concept class of multilinear formulas is
efficiently learnable.

\item[(7)] Is there a practical method to compute the tree size of, say,
$10$-qubit states? \ Such a method would have great value in interpreting
experimental results.
\end{enumerate}

\section*{Acknowledgments}

I thank Ran Raz for fruitful correspondence and for sharing an early version
of his paper; the anonymous reviewers for detailed comments that improved the
paper enormously; and Andrej Bogdanov, Don Coppersmith, Viatcheslav
Dobrovitski, Oded Goldreich, Ray Laflamme, Anthony Leggett, Leonid Levin, Mike
Mosca, Ashwin Nayak, Carl Pomerance, John Preskill, Alexander Razborov, Peter
Shor, Rob Spekkens, Barbara Terhal, Luca Trevisan, Umesh Vazirani, Guifre
Vidal, and Avi Wigderson for helpful discussions.

\section{Appendix: Relations Among Quantum State Classes\label{REL}}

This appendix presents some results about the quantum state hierarchy
introduced in Section \ref{CQS}. \ Theorem \ref{trivrelate}\ shows simple
inclusions and separations, while Theorem \ref{psipampp} shows that
separations higher in the hierarchy would imply major complexity class
separations (and vice versa).

\begin{theorem}
\label{trivrelate}\quad

\begin{enumerate}
\item[(i)] $\mathsf{Tree}\cup\mathsf{Vidal}\subseteq\mathsf{Circuit}%
\subseteq\mathsf{AmpP}$.

\item[(ii)] All states in $\mathsf{Vidal}$ have tree size $n^{O\left(  \log
n\right)  }$.

\item[(iii)] $\mathsf{\Sigma}_{\mathsf{2}}\subseteq\mathsf{Vidal}$ but
$\mathsf{\otimes}_{\mathsf{2}}\not \subset \mathsf{Vidal}$.

\item[(iv)] $\mathsf{\otimes}_{\mathsf{2}}\subsetneq\mathsf{MOTree}$.

\item[(v)] $\mathsf{\Sigma}_{\mathsf{1}}$, $\mathsf{\Sigma}_{\mathsf{2}}$,
$\mathsf{\Sigma}_{\mathsf{3}}$, $\mathsf{\otimes}_{\mathsf{1}}$,
$\mathsf{\otimes}_{\mathsf{2}}$, and $\mathsf{\otimes}_{\mathsf{3}}$ are all
distinct. \ Also, $\mathsf{\otimes}_{\mathsf{3}}\neq\mathsf{\Sigma
}_{\mathsf{4}}\cap\mathsf{\otimes}_{\mathsf{4}}$.
\end{enumerate}
\end{theorem}

\begin{proof}

\begin{enumerate}
\item[(i)] $\mathsf{Tree}\subseteq\mathsf{Circuit}$ since any multilinear tree
is also a multilinear circuit. \ $\mathsf{Circuit}\subseteq\mathsf{AmpP}%
$\ since the circuit yields a polynomial-time algorithm for computing the
amplitudes. \ For\ $\mathsf{Vidal}\subseteq\mathsf{Circuit}$, we use an idea
of Vidal \cite{vidal}: given $\left\vert \psi_{n}\right\rangle \in
\mathsf{Vidal}$, for all $j\in\left\{  1,\ldots,n\right\}  $\ we can express
$\left\vert \psi_{n}\right\rangle $\ as%
\[
\sum_{i=1}^{\chi\left(  \left\vert \psi\right\rangle \right)  }\alpha
_{ij}\left\vert \phi_{i}^{\left[  1\ldots j\right]  }\right\rangle
\otimes\left\vert \phi_{i}^{\left[  j+1\ldots n\right]  }\right\rangle
\]
where $\chi\left(  \left\vert \psi_{n}\right\rangle \right)  $\ is
polynomially bounded. \ Furthermore, Vidal showed that each $\left\vert
\phi_{i}^{\left[  1\ldots j\right]  }\right\rangle $\ can be written as a
linear combination of states of the form $\left\vert \phi_{i}^{\left[  1\ldots
j-1\right]  }\right\rangle \otimes\left\vert 0\right\rangle $\ and $\left\vert
\phi_{i}^{\left[  1\ldots j-1\right]  }\right\rangle \otimes\left\vert
1\right\rangle $---the point being that the set of $\left\vert \phi
_{i}^{\left[  1\ldots j-1\right]  }\right\rangle $\ states is the same,
independently of $\left\vert \phi_{i}^{\left[  1\ldots j\right]
}\right\rangle $. \ This immediately yields a polynomial-size multilinear
circuit for $\left\vert \psi_{n}\right\rangle $.

\item[(ii)] Given $\left\vert \psi_{n}\right\rangle \in\mathsf{Vidal}$, we can
decompose $\left\vert \psi_{n}\right\rangle $\ as%
\[
\sum_{i=1}^{\chi\left(  \left\vert \psi\right\rangle \right)  }\alpha
_{i}\left\vert \phi_{i}^{\left[  1\ldots n/2\right]  }\right\rangle
\otimes\left\vert \phi_{i}^{\left[  n/2+1\ldots n\right]  }\right\rangle .
\]
Then $\chi\left(  \left\vert \phi_{i}^{\left[  1\ldots n/2\right]
}\right\rangle \right)  \leq\chi\left(  \left\vert \psi_{n}\right\rangle
\right)  $\ and $\chi\left(  \left\vert \phi_{i}^{\left[  n/2+1\ldots
n\right]  }\right\rangle \right)  \leq\chi\left(  \left\vert \psi
_{n}\right\rangle \right)  $ for all $i$, so we can recursively decompose
these states in the same manner. \ It follows that $\operatorname*{TS}\left(
\left\vert \psi_{n}\right\rangle \right)  \leq2\chi\left(  \left\vert
\psi\right\rangle \right)  \operatorname*{TS}\left(  \left\vert \psi
_{n/2}\right\rangle \right)  $; solving this recurrence relation yields
$\operatorname*{TS}\left(  \left\vert \psi_{n}\right\rangle \right)
\leq\left(  2\chi\left(  \left\vert \psi\right\rangle \right)  \right)  ^{\log
n}=n^{O\left(  \log n\right)  }$.

\item[(iii)] $\mathsf{\Sigma}_{\mathsf{2}}\subseteq\mathsf{Vidal}$ follows
since a sum of $t$ separable states has $\chi\leq t$,\ while $\mathsf{\otimes
}_{\mathsf{2}}\not \subset \mathsf{Vidal}$\ follows from the example of $n/2$
Bell pairs: $2^{-n/4}\left(  \left\vert 00\right\rangle +\left\vert
11\right\rangle \right)  ^{\otimes n/2}$.

\item[(iv)] $\mathsf{\otimes}_{\mathsf{2}}\subseteq\mathsf{MOTree}$ is
obvious, while\ $\mathsf{MOTree}\not \subset \mathsf{\otimes}_{\mathsf{2}}%
$\ follows from the example of $\left\vert P_{n}^{i}\right\rangle $, an equal
superposition over all $n$-bit strings of parity $i$. \ The following
recursive formulas imply that $\operatorname*{MOTS}\left(  \left\vert
P_{n}^{i}\right\rangle \right)  \leq4\operatorname*{MOTS}\left(  \left\vert
P_{n/2}^{i}\right\rangle \right)  =O\left(  n^{2}\right)  $:%
\begin{align*}
\left\vert P_{n}^{0}\right\rangle  &  =\frac{1}{\sqrt{2}}\left(  \left\vert
P_{n/2}^{0}\right\rangle \left\vert P_{n/2}^{0}\right\rangle +\left\vert
P_{n/2}^{1}\right\rangle \left\vert P_{n/2}^{1}\right\rangle \right)  ,\\
\left\vert P_{n}^{1}\right\rangle  &  =\frac{1}{\sqrt{2}}\left(  \left\vert
P_{n/2}^{0}\right\rangle \left\vert P_{n/2}^{1}\right\rangle +\left\vert
P_{n/2}^{1}\right\rangle \left\vert P_{n/2}^{0}\right\rangle \right)  .
\end{align*}
On the other hand, $\left\vert P_{n}\right\rangle \notin\mathsf{\otimes
}_{\mathsf{2}}$\ follows from $\left\vert P_{n}\right\rangle \notin
\mathsf{\Sigma}_{\mathsf{1}}$\ together with the fact that $\left\vert
P_{n}\right\rangle $\ has no nontrivial tensor product decomposition.

\item[(v)] $\mathsf{\otimes}_{\mathsf{1}}\not \subset \mathsf{\Sigma
}_{\mathsf{1}}$ and $\mathsf{\Sigma}_{\mathsf{1}}\not \subset \mathsf{\otimes
}_{\mathsf{1}}\ $are obvious.$\ \ \mathsf{\otimes}_{\mathsf{2}}\not \subset
\mathsf{\Sigma}_{\mathsf{2}}$ (and hence $\mathsf{\otimes}_{\mathsf{1}}%
\neq\mathsf{\otimes}_{\mathsf{2}}$) follows from part (iii). \ $\mathsf{\Sigma
}_{\mathsf{2}}\not \subset \mathsf{\otimes}_{\mathsf{2}}$\ (and hence
$\mathsf{\Sigma}_{\mathsf{1}}\neq\mathsf{\Sigma}_{\mathsf{2}}$) follows from
part (iv), together with the fact that $\left\vert P_{n}\right\rangle $ has a
$\mathsf{\Sigma}_{\mathsf{2}}$\ formula based on the Fourier transform:%
\[
\left\vert P_{n}\right\rangle =\frac{1}{\sqrt{2}}\left(  \left(
\frac{\left\vert 0\right\rangle +\left\vert 1\right\rangle }{\sqrt{2}}\right)
^{\otimes n}+\left(  \frac{\left\vert 0\right\rangle -\left\vert
1\right\rangle }{\sqrt{2}}\right)  ^{\otimes n}\right)  .
\]
$\mathsf{\Sigma}_{\mathsf{2}}\neq\mathsf{\Sigma}_{\mathsf{3}}$ follows from
$\mathsf{\otimes}_{\mathsf{2}}\not \subset \mathsf{\Sigma}_{\mathsf{2}}$\ and
$\mathsf{\otimes}_{\mathsf{2}}\subseteq\mathsf{\Sigma}_{\mathsf{3}}$. \ Also,
$\mathsf{\Sigma}_{\mathsf{3}}\not \subset \mathsf{\otimes}_{\mathsf{3}}$
follows from $\mathsf{\Sigma}_{\mathsf{2}}\neq\mathsf{\Sigma}_{\mathsf{3}}$,
together with the fact that we can easily construct states in $\mathsf{\Sigma
}_{\mathsf{3}}\setminus\mathsf{\Sigma}_{\mathsf{2}}$\ that have no nontrivial
tensor product decomposition---for example,%
\[
\frac{1}{\sqrt{2}}\left(  \left\vert 0\right\rangle ^{\otimes n}+\left(
\frac{\left\vert 01\right\rangle +\left\vert 10\right\rangle }{\sqrt{2}%
}\right)  ^{\otimes n/2}\right)  .
\]
$\mathsf{\otimes}_{\mathsf{2}}\neq\mathsf{\otimes}_{\mathsf{3}}\ $follows from
$\mathsf{\Sigma}_{\mathsf{2}}\not \subset \mathsf{\otimes}_{\mathsf{2}}$\ and
$\mathsf{\Sigma}_{\mathsf{2}}\subseteq\mathsf{\otimes}_{\mathsf{3}}$.
\ Finally, $\mathsf{\otimes}_{\mathsf{3}}\neq\mathsf{\Sigma}_{\mathsf{4}}%
\cap\mathsf{\otimes}_{\mathsf{4}}$\ follows from $\mathsf{\Sigma}_{\mathsf{3}%
}\not \subset \mathsf{\otimes}_{\mathsf{3}}$\ and $\mathsf{\Sigma}%
_{\mathsf{3}}\subseteq\mathsf{\Sigma}_{\mathsf{4}}\cap\mathsf{\otimes
}_{\mathsf{4}}$.
\end{enumerate}
\end{proof}

\begin{theorem}
\label{psipampp}\quad

\begin{enumerate}
\item[(i)] $\mathsf{BQP}=\mathsf{P}^{\mathsf{\#P}}$ implies $\mathsf{AmpP}%
\subseteq\mathsf{\Psi P}$.

\item[(ii)] $\mathsf{AmpP}\subseteq\mathsf{\Psi P}$ implies $\mathsf{NP}%
\subseteq\mathsf{BQP/poly}.$

\item[(iii)] $\mathsf{P}=\mathsf{P}^{\mathsf{\#P}}$ implies $\mathsf{\Psi
P}\subseteq\mathsf{AmpP}$.

\item[(iv)] $\mathsf{\Psi P}\subseteq\mathsf{AmpP}$ implies $\mathsf{BQP}%
\subseteq\mathsf{P/poly}$.
\end{enumerate}
\end{theorem}

\begin{proof}

\begin{enumerate}
\item[(i)] First, $\mathsf{BQP}=\mathsf{P}^{\mathsf{\#P}}$\ implies
$\mathsf{BQP/poly}=\mathsf{P}^{\mathsf{\#P}}\mathsf{/poly}$, since given a
$\mathsf{P}^{\mathsf{\#P}}\mathsf{/poly}$\ machine $M$, the language
consisting of all $\left(  x,a\right)  $\ such that $M$ accepts on input $x$
and advice $a$ is clearly in $\mathsf{BQP}$. \ So assume $\mathsf{BQP/poly}%
=\mathsf{P}^{\mathsf{\#P}}\mathsf{/poly}$, and consider a state $\left\vert
\psi\right\rangle =\sum_{x\in\left\{  0,1\right\}  ^{n}}\alpha_{x}\left\vert
x\right\rangle $\ with\ $\left\vert \psi\right\rangle \in\mathsf{AmpP}$. \ By
the result of Bernstein and Vazirani \cite{bv}\ that $\mathsf{BQP}%
\subseteq\mathsf{P}^{\mathsf{\#P}}$, for all $b$ there exists a quantum
circuit of size polynomial in $n$ and $b$ that approximates $p_{0}=\sum
_{y\in\left\{  0,1\right\}  ^{n-1}}\left\vert \alpha_{0y}\right\vert ^{2}$, or
the probability that the first qubit is measured to be $0$, to $b$ bits of
precision. \ So by uncomputing garbage, we can prepare a state close to
$\sqrt{p_{0}}\left\vert 0\right\rangle +\sqrt{1-p_{0}}\left\vert
1\right\rangle $. \ Similarly, given a superposition over length-$k$ prefixes
of $x$, we can prepare a superposition over length-$\left(  k+1\right)  $
prefixes of $x$ by approximating the conditional measurement probabilities.
\ We thus obtain a state close to $\sum_{x}\left\vert \alpha_{x}\right\vert
\left\vert x\right\rangle $. \ The last step is to approximate the phase of
each $\left\vert x\right\rangle $, apply that phase, and uncompute to obtain a
state close to $\sum_{x}\alpha_{x}\left\vert x\right\rangle $.

\item[(ii)] Given a $SAT$ instance, first use Valiant-Vazirani \cite{vv} to
produce a formula $\varphi$\ with either $0$ or $1$ satisfying assignments.
\ Then let $\alpha_{x}=1$\ if $x$ is a satisfying assignment for $\varphi
$\ and $\alpha_{x}=0$\ otherwise; clearly $\left\vert \psi\right\rangle
=\sum_{x}\alpha_{x}\left\vert x\right\rangle $ is in $\mathsf{AmpP}$. \ By the
assumption $\mathsf{AmpP}\subseteq\mathsf{\Psi P}$, there exists a
polynomial-size quantum circuit that approximates $\left\vert \psi
\right\rangle $, and thereby finds the unique satisfying assignment for
$\varphi$\ if it exists.

\item[(iii)] As in part (i), $\mathsf{P}=\mathsf{P}^{\mathsf{\#P}}$\ implies
$\mathsf{P/poly}=\mathsf{P}^{\mathsf{\#P}}\mathsf{/poly}$. \ The containment
$\mathsf{\Psi P}\subseteq\mathsf{AmpP}$\ follows since we can approximate
amplitudes to polynomially many bits of precision in $\mathsf{\#P}$.

\item[(iv)] As is well known \cite{bv}, any quantum computation can be made
`clean' in the sense that it accepts if and only if a particular basis state
(say $\left\vert 0\right\rangle ^{\otimes n}$) is measured. \ The implication
follows easily.
\end{enumerate}
\end{proof}

\section{Appendix: Manifestly Orthogonal Tree Size\label{MOTS}}

This appendix studies the manifestly orthogonal tree size of coset
states:\footnote{All results apply equally well to the subgroup states of
Section \ref{ECC}; the greater generality of coset states is just for
convenience.} states having the form%
\[
\left\vert C\right\rangle =\frac{1}{\sqrt{\left\vert C\right\vert }}\sum_{x\in
C}\left\vert x\right\rangle
\]
where $C=\left\{  x~|~Ax\equiv b\right\}  $\ is a coset in $\mathbb{Z}_{2}%
^{n}$. \ In particular, we present a \textit{tight} characterization of
$\operatorname*{MOTS}\left(  \left\vert C\right\rangle \right)  $, which
enables us to prove \textit{exponential} lower bounds on it, in contrast to
the $n^{\Omega\left(  \log n\right)  }$ lower bounds for ordinary tree size.
\ This characterization also yields a separation between orthogonal and
manifestly orthogonal tree size; and an algorithm for computing
$\operatorname*{MOTS}\left(  \left\vert C\right\rangle \right)  $\ whose
complexity is only singly exponential in $n$. \ Our proof technique is
independent of Raz's, and is highly tailored to take advantage of manifest
orthogonality. \ However, even if our technique finds no broader application,
the fact that it gives tight bounds makes it almost unique---and thus, we
hope, of interest to complexity theorists.

Given a state $\left\vert \psi\right\rangle $, recall that the manifestly
orthogonal tree size $\operatorname*{MOTS}\left(  \left\vert \psi\right\rangle
\right)  $ is the minimum size of a tree representing $\left\vert
\psi\right\rangle $, in which all additions are of two states $\left\vert
\psi_{1}\right\rangle ,\left\vert \psi_{2}\right\rangle $\ with
\textquotedblleft disjoint supports\textquotedblright---that
is,\ either\ $\left\langle \psi_{1}|x\right\rangle =0$ or $\left\langle
\psi_{2}|x\right\rangle =0$ for every basis state $\left\vert x\right\rangle
$. \ Here the size $\left\vert T\right\vert $\ of $T$ is the number of leaf
vertices. \ We can assume without loss of generality that every $+$\ or
$\otimes$\ vertex has at least one child, and that every child of a
$+$\ vertex is a $\otimes$\ vertex and vice versa. \ Also, given a set
$S\subseteq\left\{  0,1\right\}  ^{n}$, let%
\[
\left\vert S\right\rangle =\frac{1}{\sqrt{\left\vert S\right\vert }}\sum_{x\in
S}\left\vert x\right\rangle
\]
be a uniform superposition over the elements of $S$, and let $M\left(
S\right)  $ be a shorthand for$\ \operatorname*{MOTS}\left(  \left\vert
S\right\rangle \right)  $.

Let $C=\left\{  x:Ax\equiv b\right\}  $ be a subgroup in $\mathbb{Z}_{2}^{n}$,
for some $A\in\mathbb{Z}_{2}^{k\times n}$\ and $b\in\mathbb{Z}_{2}^{k}%
$.\ \ Let $\left[  n\right]  =\left\{  1,\ldots,n\right\}  $, and let $\left(
I,J\right)  $\ be a nontrivial partition of $\left[  n\right]  $ (one where
$I$\ and $J$ are both nonempty). \ Then clearly there exist distinct cosets
$C_{I}^{\left(  1\right)  },\ldots,C_{I}^{\left(  H\right)  }$\ in the $I$
subsystem, and distinct cosets\ $C_{J}^{\left(  1\right)  },\ldots
,C_{J}^{\left(  H\right)  }$\ in the $J$\ subsystem, such that%
\[
C=%
{\displaystyle\bigcup\limits_{h\in\left[  H\right]  }}
C_{I}^{\left(  h\right)  }\otimes C_{J}^{\left(  h\right)  }.
\]
The $C_{I}^{\left(  h\right)  }$'s\ and $C_{J}^{\left(  h\right)  }$'s\ are
unique up to ordering.\ \ Furthermore, the quantities $\left\vert
C_{I}^{\left(  h\right)  }\right\vert $, $\left\vert C_{J}^{\left(  h\right)
}\right\vert $, $M\left(  C_{I}^{\left(  h\right)  }\right)  $, and $M\left(
C_{J}^{\left(  h\right)  }\right)  $\ remain unchanged as we range over
$h\in\left[  H\right]  $. \ For this reason we suppress the dependence on $h$
when mentioning them.

For various sets $S$, our strategy will be to analyze $M\left(  S\right)
/\left\vert S\right\vert $, the ratio of tree size to cardinality. \ We can
think of this ratio as the \textquotedblleft price per pound\textquotedblright%
\ of $S$: the number of vertices that we have to pay per basis state that we
cover. \ The following lemma says that, under that cost measure, a coset is
\textquotedblleft as good a deal\textquotedblright\ as any of its subsets:

\begin{lemma}
\label{induclem}For all cosets $C$,%
\[
\frac{M\left(  C\right)  }{\left\vert C\right\vert }=\min\left(
\frac{M\left(  S\right)  }{\left\vert S\right\vert }\right)
\]
where the minimum is over nonempty $S\subseteq C$.
\end{lemma}

\begin{proof}
By induction on $n$. \ The base case $n=1$\ is obvious, so\ assume the lemma
true for $n-1$. \ Choose $S^{\ast}\subseteq C$ to minimize $M\left(  S^{\ast
}\right)  /\left\vert S^{\ast}\right\vert $. \ Let $T$\ be a manifestly
orthogonal tree for $\left\vert S^{\ast}\right\rangle $ of minimum size, and
let $v$ be the root of $T$. \ We can assume without loss of generality that
$v$ is a $\otimes$ vertex, since otherwise $v$ has some $\otimes$\ child
representing a set $R\subset S^{\ast}$\ such that $M\left(  R\right)
/\left\vert R\right\vert \leq M\left(  S^{\ast}\right)  /\left\vert S^{\ast
}\right\vert $. \ Therefore for some nontrivial partition $\left(  I,J\right)
$\ of $\left[  n\right]  $, and some $S_{I}^{\ast}\subseteq\left\{
0,1\right\}  ^{\left\vert I\right\vert }$ and $S_{J}^{\ast}\subseteq\left\{
0,1\right\}  ^{\left\vert J\right\vert }$, we have%
\begin{align*}
\left\vert S^{\ast}\right\rangle  &  =\left\vert S_{I}^{\ast}\right\rangle
\otimes\left\vert S_{J}^{\ast}\right\rangle ,\\
\left\vert S^{\ast}\right\vert  &  =\left\vert S_{I}^{\ast}\right\vert
\left\vert S_{J}^{\ast}\right\vert ,\\
M\left(  S^{\ast}\right)   &  =M\left(  S_{I}^{\ast}\right)  +M\left(
S_{J}^{\ast}\right)  ,
\end{align*}
where the last equality holds because if $M\left(  S^{\ast}\right)  <M\left(
S_{I}^{\ast}\right)  +M\left(  S_{J}^{\ast}\right)  $, then $T$ was not a
minimal tree for $\left\vert S^{\ast}\right\rangle $. \ Then%
\[
\frac{M\left(  S^{\ast}\right)  }{\left\vert S^{\ast}\right\vert }%
=\frac{M\left(  S_{I}^{\ast}\right)  +M\left(  S_{J}^{\ast}\right)
}{\left\vert S_{I}^{\ast}\right\vert \left\vert S_{J}^{\ast}\right\vert }%
=\min\left(  \frac{M\left(  S_{I}\right)  +M\left(  S_{J}\right)  }{\left\vert
S_{I}\right\vert \left\vert S_{J}\right\vert }\right)
\]
where the minimum is over nonempty $S_{I}\subseteq\left\{  0,1\right\}
^{\left\vert I\right\vert }$\ and $S_{J}\subseteq\left\{  0,1\right\}
^{\left\vert J\right\vert }$\ such that $S_{I}\otimes S_{J}\subseteq C$. \ Now
there must be an $h$ such that $S_{I}^{\ast}\subseteq C_{I}^{\left(  h\right)
}$\ and $S_{J}^{\ast}\subseteq C_{J}^{\left(  h\right)  }$, since otherwise
some $x\notin C$\ would be assigned nonzero amplitude. \ By the induction
hypothesis,%
\[
\frac{M\left(  C_{I}\right)  }{\left\vert C_{I}\right\vert }=\min\left(
\frac{M\left(  S_{I}\right)  }{\left\vert S_{I}\right\vert }\right)
,~~~~~~~~~~\frac{M\left(  C_{J}\right)  }{\left\vert C_{J}\right\vert }%
=\min\left(  \frac{M\left(  S_{J}\right)  }{\left\vert S_{J}\right\vert
}\right)  ,
\]
where the minima are over nonempty $S_{I}\subseteq C_{I}^{\left(  h\right)  }$
and $S_{J}\subseteq C_{J}^{\left(  h\right)  }$\ respectively. \ Define
$\beta=\left\vert S_{I}\right\vert \cdot\left\vert S_{J}\right\vert /M\left(
S_{J}\right)  $\ and $\gamma=\left\vert S_{J}\right\vert \cdot\left\vert
S_{I}\right\vert /M\left(  S_{I}\right)  $. \ Then since setting $S_{I}%
:=C_{I}^{\left(  h\right)  }$\ and $S_{J}:=C_{J}^{\left(  h\right)  }%
$\ maximizes the four quantities $\left\vert S_{I}\right\vert $,\ $\left\vert
S_{J}\right\vert $, $\left\vert S_{I}\right\vert /M\left(  S_{I}\right)  $,
and $\left\vert S_{J}\right\vert /M\left(  S_{J}\right)  $\ simultaneously,
this choice also maximizes $\beta$\ and $\gamma$ simultaneously. \ Therefore
it maximizes their harmonic mean,%
\[
\frac{\beta\gamma}{\beta+\gamma}=\frac{\left\vert S_{I}\right\vert \left\vert
S_{J}\right\vert }{M\left(  S_{I}\right)  +M\left(  S_{J}\right)  }%
=\frac{\left\vert S\right\vert }{M\left(  S\right)  }.
\]
We have proved that setting $S:=C_{I}^{\left(  h\right)  }\otimes
C_{J}^{\left(  h\right)  }$\ maximizes $\left\vert S\right\vert /M\left(
S\right)  $, or equivalently minimizes $M\left(  S\right)  /\left\vert
S\right\vert $. \ The one remaining observation is that taking the disjoint
sum of $C_{I}^{\left(  h\right)  }\otimes C_{J}^{\left(  h\right)  }$\ over
all $h\in\left[  H\right]  $\ leaves the ratio $M\left(  S\right)  /\left\vert
S\right\vert $ unchanged. \ So setting $S:=C$\ also minimizes $M\left(
S\right)  /\left\vert S\right\vert $, and we are done.
\end{proof}

We are now ready to give a recursive characterization of $M\left(  C\right)  $.

\begin{theorem}
\label{motstight}If $n\geq2$, then%
\[
M\left(  C\right)  =\left\vert C\right\vert \min\left(  \frac{M\left(
C_{I}\right)  +M\left(  C_{J}\right)  }{\left\vert C_{I}\right\vert \left\vert
C_{J}\right\vert }\right)
\]
where the minimum is over nontrivial partitions $\left(  I,J\right)  $ of
$\left[  n\right]  $.
\end{theorem}

\begin{proof}
The upper bound is obvious; we prove the lower bound. \ Let $T$ be a
manifestly orthogonal tree for $\left\vert C\right\rangle $ of minimum size,
and let $v^{\left(  1\right)  },\ldots,v^{\left(  L\right)  }$\ be the topmost
$\otimes$\ vertices in $T$. \ Then there exists a partition $\left(
S^{\left(  1\right)  },\ldots,S^{\left(  L\right)  }\right)  $\ of $C$ such
that the subtree rooted at $v^{\left(  i\right)  }$\ represents $\left\vert
S^{\left(  i\right)  }\right\rangle $. \ We have%
\[
\left\vert T\right\vert =M\left(  S^{\left(  1\right)  }\right)
+\cdots+M\left(  S^{\left(  L\right)  }\right)  =\left\vert S^{\left(
1\right)  }\right\vert \frac{M\left(  S^{\left(  1\right)  }\right)
}{\left\vert S^{\left(  1\right)  }\right\vert }+\cdots+\left\vert S^{\left(
L\right)  }\right\vert \frac{M\left(  S^{\left(  L\right)  }\right)
}{\left\vert S^{\left(  L\right)  }\right\vert }.
\]
Now let\ $\eta=\min_{i}\left(  M\left(  S^{\left(  i\right)  }\right)
/\left\vert S^{\left(  i\right)  }\right\vert \right)  $. \ We will construct
a partition $\left(  R^{\left(  1\right)  },\ldots,R^{\left(  H\right)
}\right)  $\ of $C$ such that $M\left(  R^{\left(  h\right)  }\right)
/\left\vert R^{\left(  h\right)  }\right\vert =\eta$\ for all $h\in\left[
H\right]  $, which will imply a new tree $T^{\prime}$ with $\left\vert
T^{\prime}\right\vert \leq\left\vert T\right\vert $. \ Choose $j\in\left[
L\right]  $\ such that $M\left(  S^{\left(  j\right)  }\right)  /\left\vert
S^{\left(  j\right)  }\right\vert =\eta$, and suppose vertex $v^{\left(
j\right)  }$\ of $T$ expresses $\left\vert S^{\left(  j\right)  }\right\rangle
$ as $\left\vert S_{I}\right\rangle \otimes\left\vert S_{J}\right\rangle
$\ for some nontrivial partition $\left(  I,J\right)  $. \ Then%
\[
\eta=\frac{M\left(  S^{\left(  j\right)  }\right)  }{\left\vert S^{\left(
j\right)  }\right\vert }=\frac{M\left(  S_{I}\right)  +M\left(  S_{J}\right)
}{\left\vert S_{I}\right\vert \left\vert S_{J}\right\vert }%
\]
where $M\left(  S^{\left(  j\right)  }\right)  =M\left(  S_{I}\right)
+M\left(  S_{J}\right)  $ follows from the minimality of $T$. \ As in Lemma
\ref{induclem}, there must be an $h$ such that $S_{I}\subseteq C_{I}^{\left(
h\right)  }$\ and $S_{J}\subseteq C_{J}^{\left(  h\right)  }$. \ But Lemma
\ref{induclem}\ then implies that $M\left(  C_{I}\right)  /\left\vert
C_{I}\right\vert \leq M\left(  S_{I}\right)  /\left\vert S_{I}\right\vert
$\ and that $M\left(  C_{J}\right)  /\left\vert C_{J}\right\vert \leq M\left(
S_{J}\right)  /\left\vert S_{J}\right\vert $. \ Combining these bounds with
$\left\vert C_{I}\right\vert \geq\left\vert S_{I}\right\vert $\ and
$\left\vert C_{J}\right\vert \geq\left\vert S_{J}\right\vert $, we obtain by a
harmonic mean inequality that%
\[
\frac{M\left(  C_{I}\otimes C_{J}\right)  }{\left\vert C_{I}\otimes
C_{J}\right\vert }\leq\frac{M\left(  C_{I}\right)  +M\left(  C_{J}\right)
}{\left\vert C_{I}\right\vert \left\vert C_{J}\right\vert }\leq\frac{M\left(
S_{I}^{\ast}\right)  +M\left(  S_{J}^{\ast}\right)  }{\left\vert S_{I}^{\ast
}\right\vert \left\vert S_{J}^{\ast}\right\vert }=\eta.
\]
So setting $R^{\left(  h\right)  }:=C_{I}^{\left(  h\right)  }\otimes
C_{J}^{\left(  h\right)  }$\ for all $h\in\left[  H\right]  $\ yields a new
tree $T^{\prime}$\ no larger than $T$. \ Hence by the minimality of $T$,%
\[
M\left(  C\right)  =\left\vert T\right\vert =\left\vert T^{\prime}\right\vert
=H\cdot M\left(  C_{I}\otimes C_{J}\right)  =\frac{\left\vert C\right\vert
}{\left\vert C_{I}\right\vert \left\vert C_{J}\right\vert }\cdot\left(
M\left(  C_{I}\right)  +M\left(  C_{J}\right)  \right)  .
\]

\end{proof}

We can express Theorem\ \ref{motstight}\ directly in terms of the matrix $A$
as follows. \ Let $M\left(  A\right)  =M\left(  C\right)
=\operatorname*{MOTS}\left(  \left\vert C\right\rangle \right)  $\ where
$C=\left\{  x:Ax\equiv b\right\}  $\ (the vector $b$ is irrelevant, so long as
$Ax\equiv b$\ is solvable). \ Then%
\begin{equation}
M\left(  A\right)  =\min\left(  2^{\operatorname*{rank}\left(  A_{I}\right)
+\operatorname*{rank}\left(  A_{J}\right)  -\operatorname*{rank}\left(
A\right)  }\left(  M\left(  A_{I}\right)  +M\left(  A_{J}\right)  \right)
\right)  \tag{*}\label{marank}%
\end{equation}
where the minimum is over all nontrivial partitions $\left(  A_{I}%
,A_{J}\right)  $\ of the columns of $A$. \ As a base case, if $A$ has only one
column, then $M\left(  A\right)  =2$ if $A=0$\ and $M\left(  A\right)
=1$\ otherwise. \ This immediately implies the following.

\begin{corollary}
\label{algcor}There exists a deterministic $O\left(  n3^{n}\right)  $-time
algorithm that computes $M\left(  A\right)  $, given $A$ as input.
\end{corollary}

\begin{proof}
First compute $\operatorname*{rank}\left(  A^{\ast}\right)  $\ for all
$2^{n-1}$\ matrices $A^{\ast}$\ that are formed by choosing a subset of the
columns of $A$. \ This takes time $O\left(  n^{3}2^{n}\right)  $.\ \ Then
compute $M\left(  A^{\ast}\right)  $\ for all $A^{\ast}$\ with one column,
then for all $A^{\ast}$\ with two columns, and so on, applying the formula
(\ref{marank})\ recursively. \ This takes time%
\[
\sum_{t=1}^{n}\dbinom{n}{t}t2^{t}=O\left(  n3^{n}\right)  .
\]

\end{proof}

Another easy consequence of Theorem \ref{motstight} is that the language
$\left\{  A:M\left(  A\right)  \leq s\right\}  $\ is in $\mathsf{NP}$. \ We do
not know whether this language is $\mathsf{NP}$-complete but suspect it is.

As we mentioned, our characterization lets us prove exponential lower bounds
on the manifestly orthogonal tree size of coset states.

\begin{theorem}
\label{explb}Suppose the entries of $A\in\mathbb{Z}_{2}^{k\times n}$ are drawn
uniformly and independently at random, where $k\in\left[  4\log_{2}n,\frac
{1}{2}\sqrt{n\ln2}\right]  $. \ Then $M\left(  A\right)  $\ $=\left(
n/k^{2}\right)  ^{\Omega\left(  k\right)  }$\ with probability $\Omega\left(
1\right)  $ over $A$.
\end{theorem}

\begin{proof}
Let us upper-bound the probability that certain \textquotedblleft bad
events\textquotedblright\ occur when $A$ is drawn. \ The first bad event is
that $A$ contains an all-zero column. \ This occurs with probability at most
$2^{-k}n=o\left(  1\right)  $. \ The second bad event is that there exists a
$k\times d$\ submatrix of $A$ with $d\geq12k$\ that has rank at most $2k/3$.
\ This also occurs with probability $o\left(  1\right)  $. \ For we claim
that, if $A^{\ast}$\ is drawn uniformly at random from $\mathbb{Z}%
_{2}^{k\times d}$, then%
\[
\Pr_{A_{I}}\left[  \operatorname*{rank}\left(  A^{\ast}\right)  \leq r\right]
\leq\dbinom{d}{r}\left(  \frac{2^{r}}{2^{k}}\right)  ^{d-r}.
\]
To see this, imagine choosing the columns of $A^{\ast}$\ one by one. \ For
$\operatorname*{rank}\left(  A^{\ast}\right)  $ to be at most $r$, there must
be at least $d-r$\ columns that are linearly dependent on the previous
columns. \ But each column is dependent on the previous ones with probability
at most $2^{r}/2^{k}$. \ The claim then follows from the union bound. \ So the
probability that \textit{any} $k\times d$\ submatrix of $A$ has rank at most
$r$ is at most%
\[
\dbinom{n}{d}\dbinom{d}{r}\left(  \frac{2^{r}}{2^{k}}\right)  ^{d-r}\leq
n^{d}d^{r}\left(  \frac{2^{r}}{2^{k}}\right)  ^{d-r}.
\]
Set $r=2k/3$ and $d=12k$; then the above is at most%
\[
\exp\left\{  12k\log n+\frac{2k}{3}\log\left(  12k\right)  -\left(
12k-\frac{2k}{3}\right)  \frac{k}{3}\right\}  =o\left(  1\right)
\]
where we have used the fact that $k\geq4\log n$.

Assume that neither bad event occurs, and let $\left(  A_{I}^{\left(
0\right)  },A_{J}^{\left(  0\right)  }\right)  $ be a partition of the columns
of $A$ that minimizes the expression (\ref{marank}). \ Let $A^{\left(
1\right)  }=A_{I}^{\left(  0\right)  }$\ if $\left\vert A_{I}^{\left(
0\right)  }\right\vert \geq\left\vert A_{J}^{\left(  0\right)  }\right\vert
$\ and $A^{\left(  1\right)  }=A_{J}^{\left(  0\right)  }$\ otherwise, where
$\left\vert A_{I}^{\left(  0\right)  }\right\vert $\ and $\left\vert
A_{J}^{\left(  0\right)  }\right\vert $\ are the numbers of columns in
$A_{I}^{\left(  0\right)  }$\ and $A_{J}^{\left(  0\right)  }$\ respectively
(so that $\left\vert A_{I}^{\left(  0\right)  }\right\vert +\left\vert
A_{J}^{\left(  0\right)  }\right\vert =n$). \ Likewise, let $\left(
A_{I}^{\left(  1\right)  },A_{J}^{\left(  1\right)  }\right)  $\ be an optimal
partition of the columns of $A^{\left(  1\right)  }$, and let $A^{\left(
2\right)  }=A_{I}^{\left(  1\right)  }$\ if $\left\vert A_{I}^{\left(
1\right)  }\right\vert \geq\left\vert A_{J}^{\left(  1\right)  }\right\vert
$\ and $A^{\left(  2\right)  }=A_{J}^{\left(  1\right)  }$\ otherwise.
\ Continue in this way until an $A^{\left(  t\right)  }$\ is reached such that
$\left\vert A^{\left(  t\right)  }\right\vert =1$. \ Then an immediate
consequence of (\ref{marank})\ is that $M\left(  A\right)  \geq Z^{\left(
0\right)  }\cdot\cdots\cdot Z^{\left(  t-1\right)  }$\ where%
\[
Z^{\left(  l\right)  }=2^{\operatorname*{rank}\left(  A_{I}^{\left(  l\right)
}\right)  +\operatorname*{rank}\left(  A_{J}^{\left(  l\right)  }\right)
-\operatorname*{rank}\left(  A^{\left(  l\right)  }\right)  }%
\]
and $A^{\left(  0\right)  }=A$.

Call $l$ a \textquotedblleft balanced cut\textquotedblright\ if $\min\left\{
\left\vert A_{I}^{\left(  l\right)  }\right\vert ,\left\vert A_{J}^{\left(
l\right)  }\right\vert \right\}  \geq12k$, and an \textquotedblleft unbalanced
cut\textquotedblright\ otherwise. \ If $l$ is a balanced cut, then
$\operatorname*{rank}\left(  A_{I}^{\left(  l\right)  }\right)  \geq2k/3$\ and
$\operatorname*{rank}\left(  A_{J}^{\left(  l\right)  }\right)  \geq2k/3$, so
$Z^{\left(  l\right)  }\geq2^{k/3}$. \ If $l$ is an unbalanced cut, then call
$l$ a \textquotedblleft freebie\textquotedblright\ if $\operatorname*{rank}%
\left(  A_{I}^{\left(  l\right)  }\right)  +\operatorname*{rank}\left(
A_{J}^{\left(  l\right)  }\right)  =\operatorname*{rank}\left(  A^{\left(
l\right)  }\right)  $. \ There can be at most $k$ freebies, since for each
one, $\operatorname*{rank}\left(  A^{\left(  l+1\right)  }\right)
<\operatorname*{rank}\left(  A^{\left(  l\right)  }\right)  $\ by the
assumption that all columns of $A$\ are nonzero. \ For the other unbalanced
cuts, $Z^{\left(  l\right)  }\geq2$.

Assume $\left\vert A^{\left(  l+1\right)  }\right\vert =\left\vert A^{\left(
l\right)  }\right\vert /2$\ for each balanced cut and $\left\vert A^{\left(
l+1\right)  }\right\vert =\left\vert A^{\left(  l\right)  }\right\vert
-12k$\ for each unbalanced cut. \ Then if our goal is to minimize $Z^{\left(
0\right)  }\cdot\cdots\cdot Z^{\left(  t-1\right)  }$, clearly the best
strategy is to perform balanced cuts first, then unbalanced cuts until
$\left\vert A^{\left(  l\right)  }\right\vert =12k^{2}$, at which point we can
use the $k$ freebies. \ Let $B$ be the number of balanced cuts; then%
\[
Z^{\left(  0\right)  }\cdot\cdots\cdot Z^{\left(  t-1\right)  }=\left(
2^{k/3}\right)  ^{B}2^{\left(  n/2^{B}-12k^{2}\right)  /12k}.
\]
This is minimized by taking $B=\log_{2}\left(  \frac{n\ln2}{4k^{2}}\right)  $,
in which case $Z^{\left(  0\right)  }\cdot\cdots\cdot Z^{\left(  t-1\right)
}=\left(  n/k^{2}\right)  ^{\Omega\left(  k\right)  }$.
\end{proof}

A final application of our characterization is to separate orthogonal from
manifestly orthogonal tree size.

\begin{corollary}
\label{otreesep}There exist states with polynomially-bounded orthogonal tree
size, but manifestly orthogonal tree size $n^{\Omega\left(  \log n\right)  }$.
\ Thus $\mathsf{OTree}\neq\mathsf{MOTree}$.
\end{corollary}

\begin{proof}
Set $k=4\log_{2}n$, and let $C=\left\{  x:Ax\equiv0\right\}  $\ where $A$\ is
drawn uniformly at random from $\mathbb{Z}_{2}^{k\times n}$. \ Then by Theorem
\ref{explb},%
\[
\operatorname*{MOTS}\left(  \left\vert C\right\rangle \right)  =\left(
n/k^{2}\right)  ^{\Omega\left(  k\right)  }=n^{\Omega\left(  \log n\right)  }%
\]
with probability $\Omega\left(  1\right)  $ over $A$. \ On the other hand, if
we view $\left\vert C\right\rangle $\ in the Fourier basis (that is, apply a
Hadamard to every qubit), then the resulting state has only $2^{k}=n^{16}%
$\ basis states with nonzero amplitude, and hence has orthogonal tree size at
most $n^{17}$. \ So by Proposition \ref{invariant}, part (i),
$\operatorname*{OTS}\left(  \left\vert C\right\rangle \right)  \leq2n^{17}$ as well.
\end{proof}

Indeed, the orthogonal tree states of Corollary \ref{otreesep}\ are
superpositions over polynomially many separable states, so we also obtain that
$\mathsf{\Sigma}_{\mathsf{2}}\not \subset \mathsf{MOTree}$.

\end{document}